\documentclass[10pt,a4paper]{article}

\usepackage{amssymb}
\usepackage{amsfonts}
\usepackage{calrsfs}
\usepackage{amsmath}
\usepackage[colorlinks=true,linkcolor=red,citecolor=blue,urlcolor=black,pdfpagemode=UseNone,pdfstartview=FitH]{hyperref}
\usepackage[pdftex]{graphicx}
\usepackage{a4}
\usepackage[margin=1.25in]{geometry}
\usepackage{soul}
\usepackage{booktabs}

\usepackage{caption}		
\usepackage{subcaption}
\usepackage{xargs}                      
\usepackage[pdftex,dvipsnames]{xcolor}  

\newcommand{\ee}{\end{eqnarray*}}

\title{Estimating the impact of the COVID-19\\ pandemic using granular mortality data}
\author{Frank van Berkum\thanks{Corresponding author. University of Amsterdam, Dutch Actuarial Association's Working Group for  Mortality Research  \& PwC the Netherlands, {\tt f.vanberkum@uva.nl} .}, \ \ Bertrand Melenberg\thanks{Tilburg University  \& Dutch Actuarial Association's Committee for Mortality Research, {\tt b.melenberg@tilburguniversity.edu}.} \ \ \& \ Michel Vellekoop\thanks{University of Amsterdam \& Dutch Actuarial Association's Committee for Mortality Research,  {\tt m.h.vellekoop@uva.nl} .}}

\usepackage{fancyhdr}
\DeclareMathAlphabet{\mathpzc}{OT1}{pzc}{m}{it}
\newcommand{\x}{\mathpzc{x}}

\usepackage{marginnote}

\usepackage[colorinlistoftodos,prependcaption,textsize=tiny]{todonotes}
\newcommandx{\unsure}[2][1=]{\todo[linecolor=red,backgroundcolor=red!25,bordercolor=red,#1]{#2}}
\newcommandx{\change}[2][1=]{\todo[linecolor=blue,backgroundcolor=blue!25,bordercolor=blue,#1]{#2}}
\newcommandx{\info}[2][1=]{\todo[linecolor=OliveGreen,backgroundcolor=OliveGreen!25,bordercolor=OliveGreen,#1]{#2}}
\newcommandx{\improvement}[2][1=]{\todo[linecolor=Plum,backgroundcolor=Plum!25,bordercolor=Plum,#1]{#2}}

\begin{document}

\maketitle

\begin{abstract}
We present {an extension of the} 
 Li and Lee model to quantify mortality in five European countries during the COVID-19 pandemic. The first two factors are used to model the pre-COVID mortality, with the first layer modelling the common trend and the second layer the country-specific deviation from the common trend. We add a third layer to capture the country-specific impact of COVID-19 in 2020 and 2021 in excess of the pre-COVID trend. We use weekly mortality data from the Short Term Mortality Fluctuations Database to calibrate this third factor, and we use a more granular dataset for deaths in the Netherlands to assess the added value of more detailed data. {We use our framework to define} mortality forecasts based on different possible scenarios for {the future course the pandemic.}

\end{abstract}

 \medskip
 
 \section{Introduction and motivation}

 
In this paper we use a three-layer Li and Lee model \cite{LiLee2005} to quantify mortality in five European countries during the COVID-19 pandemic, namely, Belgium, France, Germany, Great Britain, and the Netherlands. The first two layers model the pre-COVID mortality using annual data, where the first layer quantifies the common trend in the five countries and the second layer the country-specific deviation from the common trend. Our model adds a third layer to capture the country-specific impact of the new cause of death COVID-19 in the years 2020 and 2021 in excess of the pre-COVID trend. To model and quantify this third {factor}, we make use of weekly data that have become available in the Short Term Mortality Fluctuations (STMF) dataset, which forms a part of the Human Mortality Database (HMDB) \cite{HMDB}. We annualize the weekly outcomes to generate annual mortality forecasts. We supplement our model with possible scenarios for mortality rate predictions taking into account the impact of COVID-19. 
 
The Li-Lee model is an extension of the Lee-Carter model \cite{LC1992}. In the Lee-Carter model a factor which determines the impact {of mortality changes} per age is estimated, and a time-dependent factor that describes the development over time when averaged among all ages, which therefore takes the form of a time series.  Next to the Li-Lee model \cite{LiLee2005}, which extends the Lee-Carter model by adding extra layers, many other modifications have been proposed, 
introducing, for example, {additional} factors \cite{CBD2006,Plat2009}, cohort effects \cite{RH2003}, or a more appropriate model for measurement noise \cite{Brouhns2002}.  What these and alternative models have in common (see  \cite{CBD2009} for a good overview) is the goal of an improved description of the development of survival probabilities over time for different age groups, since this allows better forecasts for all sorts of statistics for human survival and better pricing and risk management models for financial products that depend on it.
  
The calibration methods for parameters in statistical models such as the Lee-Carter model rely on historical data and these are usually available in the form of yearly observations of deaths and population numbers or related statistics. {We use annual data from 1970 up to and including 2019, retrieved from the HMDB, to calibrate the first two factors of the Li-Lee model, i.e., the pre-COVID mortality.}
However, {since COVID-19 only plays a role in the years 2020 and 2021, we use weekly data to model the third layer of the Li-Lee model. The} transition from yearly to weekly data means that some adjustments in the model need to be made.

Looking at more granular data necessitates that seasonal fluctuations in deaths during the year are taken into account since the mere assumption that deaths will be distributed over the different weeks of the year in a uniform manner can already be refuted after a first casual glance at the data. But if the actual distribution {of mortality} over weeks can be reliably deduced from historical data, and a calibrated mortality model is available for observations before the start of the pandemic, then weekly observations of deaths and exposures during the pandemic should allow us to assess its impact in terms of a time series describing its severity and a factor per age-group which determines how much its members will be affected.
  
We perform our analysis for the two genders separately, since it is well-known by now  that gender is an important risk factor for death due to COVID-19. The risk for men is higher than for women, while at the same time women may be more prone to  what has become known as the ``long COVID'' form of the disease, which can cause debilitating symptoms for a very long time period but does not seem to be fatal. Such a difference between the genders is partially explained by the distinct reactions of the immune systems which have been observed for the body's response to infection with the SARS-CoV-2 virus \cite{Takahashi}. 

Apart from age and gender, we will not consider other characteristics which may influence an individual's change in survival probabilities as a result of the SARS-CoV-2 virus.\footnote{Other risk factors, which can be medical, socioeconomic or pertaining to lifestyle choices, have been shown to increase the risk of a fatal outcome after an infection. In particular, cardiovascular disease \cite{Singh}, autoimmune diseases \cite{Elliott}, diabetes \cite{McGumaghan}, and obesity and increased blood pressure \cite{SpencerGold} appear to have a significant impact \cite{Clark}. In addition, a study of COVID-19 deaths in British hospitals found a sharp increase in the risk for people with high scores on a measurement scale for deprivation, a variable that characterizes negative socioeconomic factors such as poverty, lack of social contacts and a lower level of education. Only part of this effect could be explained by the more common occurrence of existing medical conditions that are known to increase the risk to become severely ill as a result of COVID-19 \cite{Bhaskaran}. Data from the UK also shows that the risk is not the same for people with different ethnicities: in a study that explicitly corrected for age, gender, preexisting conditions and socioeconomic factors, people with a white skin were found to have a lower risk of COVID-19 mortality than those with a different skin colour. A relatively recent large cohort study that took many of these different factors into account found that among the non-medical indicators, old age, male sex and black skin colour are the most severe ones \cite{Elliott}.\label{fn:1}}
{We also} will not make use of data that try to measure the number of deaths with known cause of death COVID-19 either. The actual number of deaths that are directly related to getting the disease will far outnumber those that have been confirmed by {administrative and other data \cite{Adam2022}}. But more importantly, our aim is also to include indirect effects on mortality as a result of the pandemic as well.\footnote{Such as the deferred care mentioned in footnote \ref{fn:1} for other conditions and the reduction in the number of deaths due to the flu or traffic accidents.}

{The two-layer Li-Lee model that we use to quantify the pre-COVID mortality shows in the first layer a clear common trend. The country-specific deviations of the common trend in the second layer {turn out not} to be stationary. Instead, we model these deviations, like the first  {factor}, using random walks with drifts. These drift terms are not statistically significantly different from zero. Therefore, we set these drift terms equal to zero in the mortality projections.} 

{To quantify the {factors which extend the traditional Li-Lee model}, we do not only need death counts, but also exposures at a weekly frequency. These exposures are not available, so we have to determine these ourselves. Given the available data, we can only approximate their values. Therefore, as a comparison, we also briefly introduce the impact of COVID-19 using a compositional data (CoDa) analysis. Such a CoDa analysis only requires death counts, but not the corresponding exposures. We conduct this CoDa analysis using Dutch weekly death counts per individual age in the years 2020 and 2021, which were provided by Statistics Netherlands. Focusing on the years 2020 and 2021, the outcomes of the CoDa analysis turn out to be quite similar to the outcomes based on the three-layer Li-Lee model.}

{The weekly data in the Short Term Mortality Fluctuations (STMF) dataset only includes the death counts over five year age ranges. We use the Dutch weekly death counts per individual age in combination with the corresponding approximated exposures to investigate the impact of using five year age groups compared to individual ages. Comparing the resulting outcomes shows that the time trends are quite similar, but the age effects, after aggregating to five year intervals, might be quite sensitive to possible cohort effects. This applies in particular to ages that correspond to the baby boom generation, borne after the second world war.}

After annualizing the weekly outcomes, we use our results to generate mortality forecasts. Since the impact of COVID-19 on future mortality is quite uncertain at this stage, we present these mortality forecasts for different scenarios, where each scenario represents a possible future evolution of COVID-19.

{ Our approach differs from related research.
For example, Robben et al. \cite{Robbenetal2022} and  Schn\"urch et al. \cite{Schnurchetal2022} 
 mention that the age profile of the impact may be different from the age profile for other causes of death, but this profile is not estimated.}
 { In the first paper, the}
 { maximum likelihood estimation}
 { problem for the time series of mortality dynamics from \cite{AG2020} is} 
 { simply  modified by giving less weight  in the likelihood function to the observation years during the pandemic. }
 { By varying the corresponding weighting parameter and by adjusting the starting values of mortality projections using a modified version of the approach of Lee and Miller \cite{LeeMiller2001}, different projections for the future forces of mortality in Belgium are generated and the effect on life expectancies is determined. Schn\"urch et al. \cite{Schnurchetal2022} provide a comparative analysis of the extra deaths in 2020 using Lee-Carter models for different countries, and Cairns-Blake Dowd models for robustness checks. The authors do not introduce a new age-dependent factor but focus on the change in the time series for mortality; their approach can therefore be interpreted as an extension of the approach by Chen \& Cox \cite{ChenCox2009}, where transitory jumps in mortality time series are assumed.

Two papers in which a distinctive age pattern is addressed explicitly are Liu and Li \cite{LiuLi2015} and Zhou and Li \cite{ZhouLi2022}. In the first paper, the consequences of an age effect of a sudden mortality shock are analyzed, but under the assumption that the shock has been observed in the past and that it only affects the year in which it occurred. In  \cite{ZhouLi2022} the  Lee-Carter model is extended by a new age-and-time effect for COVID-19 and a new time series representing the overall impact over all ages.  Parameters are estimated using a penalized quasi-likelihood maximization. The estimated age distribution of the effect of COVID-19 follows the shape of the pre-COVID distribution for mortality changes. This is even true for higher ages, possibly because it is hard to separate the effect of the pandemic from ordinary mortality changes in this approach.

The calibration results of the three-layer Li-Lee model for the five countries that we include in our study show a different pattern. 
Different age groups are affected differently in the different countries, but in all cases we find and increasing trend in age and negligible effects for the youngest ages. This is in contrast to the results reported in \cite{ZhouLi2022} but in line with studies in the epidemiological literature, such as the infection fatality ratios reported in the metastudy of Driscoll et al. \cite{Driscolletal2021}. We believe that this shows the merit of the adjustments to existing estimation methods which we propose in this paper: using a Li-Lee model for different countries to identify a common pre-pandemic trend, using weekly data during the pandemic to estimate the age distribution of its impact, and using robustness checks based a more granular dataset for individual ages and a Compositional Data analysis to verify that our extrapolation method for unknown exposures is sound.
 }

The structure of the {remainder of this} paper is as follows.
Section~\ref{sec:ModelSpec} introduces the model that will be used to analyse weekly mortality observations. 
This section describes how pre-COVID mortality assumptions are derived, the estimation of seasonal effects in mortality, and how COVID-19 age and week effects are calibrated.
Section~\ref{sec:Results} first presents results using Compositional Data analysis for which only death counts are needed. 
Then, using granular Dutch mortality death and exposure information, results are shown for the COVID-19 age and week effects (including various sensitivities), and a comparison is made between the COVID-19 effects in the Netherlands, Belgium, France, Germany and Great Britain. 
Section~\ref{sec:Forecasting} illustrates how estimated COVID-19 age and week effects can be used to construct scenarios for mortality rate predictions that are adjusted for the impact of COVID-19.
Finally, Section~\ref{sec:Conclusion} concludes.

\section{Model specification and calibration}\label{sec:ModelSpec}

In this section, we describe the framework for estimating the impact of COVID-19 on the level of mortality. We start with introducing our COVID-19 mortality model as an extension of the Li-Lee model for weekly observations to which an additional term is added to capture the impact of COVID-19. Then, we describe how the baseline level of mortality is calibrated using the usual Li-Lee model, and we investigate seasonal patterns in recent weekly mortality observations. Finally, we describe how weekly exposures and death counts are obtained and how the full model is calibrated.

\subsection{A COVID-extension for the Li-Lee model}

To assess the impact of the pandemic, we propose a model in which we distinguish a baseline specification for mortality in a country prior to the pandemic, an adjustment for seasonal effects when we make the transition from yearly to weekly data at the start of 2020, and a new age-dependent factor and time series to capture the effect of the pandemic. We thus assume that the logarithm of the force of mortality in country c for age-group\footnote{
Note that $\x$ may refer to an individual age $\x=\{x\}$ or a group of ages $\x=\{x_1,x_2,...,x_n\}$.
} $\x$ and gender g in week $w$ of year $t$ has the following structure:
\begin{equation}\label{eq:1}
\ln{\mu_{\x,\, t,w}^{\rm c,g}}=\ln{\mu_{\x,t}^{\rm c,g}}+\ln\phi_{\x, w}^{\rm c,g}+\mathfrak{B}_\x^{\rm c,g}\mathfrak{K}_{t,w}^{\rm c,g}.
\end{equation}
The first term, $\mu_{\x,t}^{\rm c,g}$, is the force of mortality for the year $t$ in a (two-layer)  Li-Lee model \cite{LiLee2005}
\begin{eqnarray}\label{eq:2}
\ln{\mu_{\x,\, t}^{\rm c, g}} &=& B_\x^{\rm g}K_{t}^{\rm g}+\alpha_\x^{\rm c,g}+\beta_\x^{\rm c,g}\kappa_{t}^{\rm c,g}.
\end{eqnarray}
This force of mortality combines a Lee-Carter specification in the first term, which forces of mortality for all countries in our chosen peer group have in common, with the last two terms that define a country-specific deviation from the common dynamics. The age-dependent parameters $B_\x^{\rm g}$, $\alpha_\x^{\rm c,g}$ and $\beta_\x^{\rm c,g}$ and the time series $K_{t}^{\rm g}$ and $\kappa_{t}^{\rm c,g}$ are calibrated using yearly historical data for time periods before the pandemic; the precise procedure will be discussed in the next subsection. The forecast values for $t=2020$ and $t=2021$ based on these parameters determine the baseline values for pre-pandemic mortality ${\mu_{\x,t}^{\rm c,g}}$ in those years since these are then based on historical data before COVID-19 had any impact. 

In Equation \eqref{eq:1} for the force of mortality per week, we add two more terms. The first, $\ln\phi_{\x, w}^{\rm c,g}$, is introduced because mortality is not evenly distributed over the different weeks of the year: there is usually more mortality in the cold winter months and less mortality during the milder months. The quantity $\phi_{\x, w}^{\rm c,g}$ describes this fluctuation of mortality over the different weeks and will be called the {\sl seasonal effect}.  The last term in \eqref{eq:1}, the product $\mathfrak{B}_\x^{\rm c,g}\mathfrak{K}_{t,w}^{\rm c,g}$, {is the third layer of our Li-Lee model. It} combines a new age effect $\mathfrak{B}_\x^{\rm c,g}$ with a new time effect $\mathfrak{K}_{t,w}^{\rm c,g}$ which is 0 for $t\le 2019$. Our specification thus preserves the model structure of a Lee-Carter or Li-Lee model, while making it possible to work with a finer dataset of weekly instead of yearly data.
     
The age effect $\mathfrak{B}_\x^{\rm c,g}$ is expected to be very different from the values that are found for $B_\x^{\rm g}$ and $\beta_\x^{\rm c,g}$ because we know that excess mortality in 2020 and 2021 was largest among the highest age-groups. For this reason, it is less accurate to describe the effect of COVID-19 by only making an adjustment in the time series $K_{t}^{\rm g}$ and $\kappa_{t}^{\rm c,g}$ for $t=2020$ and $t=2021$ while retaining the existing pre-pandemic model structure. 

Note that we do not include a term that depends only on the age and the week (similar to the terms $A_{x}^{\rm g}$ and $\alpha_{x}^{\rm c,g}$), because this would mean that we make the a priori assumption that there {could}
be a lasting effect of the virus, even if values of the corresponding time series $\mathfrak{K}_{t,w}^{\rm c,g}$ would converge to zero in the future.

When analyzing mortality on an annual basis it is common to assume the force of mortality $\mu_{x,t}$ to remain constant during the year. Since we consider mortality on a weekly basis, it may seem more appropriate to assume the force of mortality $\mu_{x,t,w}$ to gradually move from $\mu_{x,t,1}=\mu_{x,t}$ to $\mu_{x,t,w_t}=\mu_{x,t+1}$ where $w_t$ equals the last week in year $t$. However, in Section~\ref{sec:Forecasting} we show how the impact of COVID-19 on the level of mortality can be incorporated in mortality forecasts. For this purpose, it turns out to be more convenient to assume that the baseline level of mortality remains constant during the year.

\subsection{Estimation of the baseline}

The baseline mortality model  in \eqref{eq:2} contains common age-dependent parameters $B_\x^{\rm g}$, country-specific age-dependent parameters $\alpha_\x^{\rm c,g}$ and $\beta_\x^{\rm c,g}$, and the time series for the common trend $K_{t}^{\rm g}$ and country-specific deviations from the trend $\kappa_{t}^{\rm c,g}$. We use maximum likelihood methods to calibrate these parameters for individual ages, i.e., with $\x=\{x\}$, for the time period before the pandemic started. The Human Mortality Database contains historical data in terms of deaths ($D_{x,t}^{\rm c,g}$) and exposures ($E_{x,t}^{\rm c,g}$) for individual ages for our peer-group of countries, which consist of the Netherlands, Belgium, Germany, France, and the United Kingdom of Great Britain and Northern Ireland, which we will abbreviate in the country set  ${\cal C}=\rm \{NLD,BEL,DEU,FRA,GBR\}$.  

 \begin{figure}[t!]
\centerline{\includegraphics[width=\textwidth]{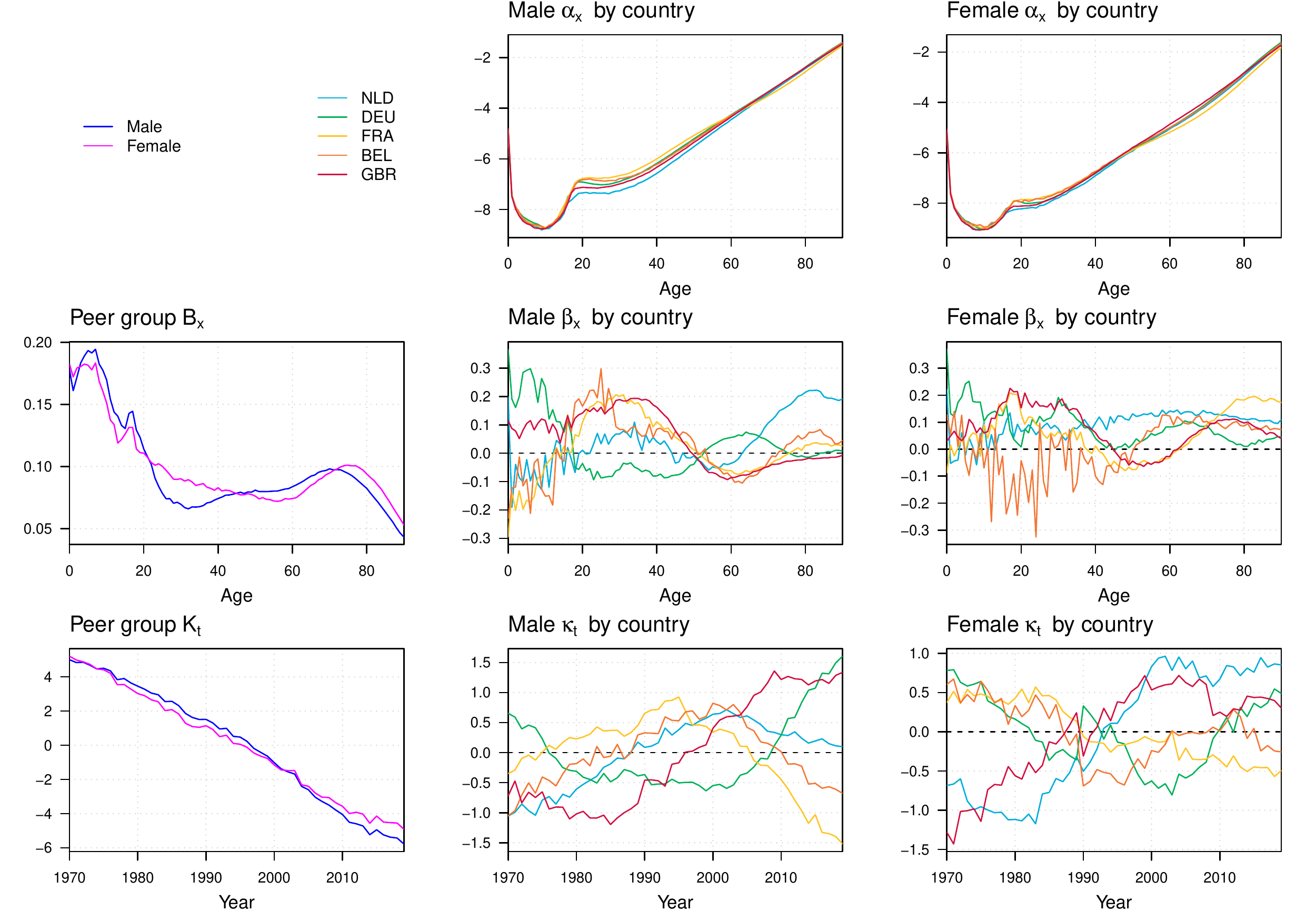}}
\caption{Estimated parameter values for the baseline model in \eqref{eq:2}. The left column shows $B_x^{\rm g}$ and $K_{t}^{\rm g}$ on the middle and the bottom row respectively. The middle column shows $\alpha_x^{\rm c,g}$, $\beta_x^{\rm c,g}$ and $\kappa_{t}^{\rm c,g}$ on the top, middle and bottom row respectively when the gender $\rm g$ is equal to male, the third column shows the same information as the second column but for females.}
\label{fig:LiLee} 
\end{figure}

Data from the years $t\in {\cal T}^{\rm EU}=\{1970,...,2019\}$ and ages $x\in {\cal X}=\{0,...,90\}$ are used for the calibration and we determine the relevant parameters in two separate steps. We first estimate the parameters which are common to all countries, $B_\x^{\rm g}$ and $K_t^{\rm g}$, using the aggregated deaths and exposures
$$
D_{x,t}^{\rm g}=\sum_{ {\rm c}\in{\cal C} }  D_{x,t}^{\rm c,g},\qquad\qquad
E_{x,t}^{\rm g}=\sum_{ {\rm c}\in{\cal C} }  E_{x,t}^{\rm c,g}.
$$
Under the distributional assumption that deaths conditioned on exposures follow the Poisson distribution
\begin{eqnarray}
D_{x,t}^{\rm g}&\sim& {\rm Poisson}\left(E_{x,t}^{\rm g} \, \mu_{x,t}^{\rm comm,g}  \right) \\
\ln  \mu_{x,t}^{\rm comm,g} &=&  A_x^{\rm g} +  B_x^{\rm g}K_{t}^{\rm g}
\end{eqnarray}
we can find parameter estimates $(A_x^{\rm g},B_x^{\rm g},K_{t}^{\rm g})$ by maximizing the log-likelihood
\begin{equation}\label{eq:3a}
	\ln L^{\rm comm} \ =  \ \sum_{ {\rm g}\in \{ {\rm m,f} \} } \sum_{x\in{\cal X}}\sum_{t\in{\cal T}} \left( D_{x,t}^{\rm g}
	 ( A_x^{\rm g} +B_x^{\rm g}K_{t}^{\rm g}) - E_{x,t}^{\rm g} \, e^{ A_x^{\rm g} +B_x^{\rm g}K_{t}^{\rm g} } \right){\ + \ C}
\end{equation}
with $C$ a constant which does not affect the optimization of the likelihood.
Once the estimates $B_x^{\rm g}$ and $K_{t}^{\rm g}$ have been determined (as well as $A_x^{\rm g}$, which will no longer be needed in the sequel), we can start the second stage of the calibration based on the specification
$$
	D_{x,t}^{\rm c,g}\sim {\rm Poisson}\left(E_{x,t}^{\rm c,g} \, \mu_{x,t}^{\rm c,g}  \right),
$$
with $\mu_{x,t}^{\rm c,g}$ as defined in \eqref{eq:2} with $\x=\{x\}$. This implies that the remaining parameters $\alpha_x^{\rm c,g}$, $\beta_x^{\rm c,g}$ and $\kappa_{t}^{\rm c,g}$ can be determined by maximization of the log-likelihood
\begin{equation}\label{eq:3}
	\ln L \ = \ \sum_{ {\rm g}\in \{ {\rm m,f} \} }\sum_{ {\rm c}\in{\cal C} } \sum_{x\in{\cal X}}\sum_{t\in{\cal T}} \left( D_{x,t}^{\rm c,g}
	 (\alpha_x^{\rm c,g}+\beta_x^{\rm c,g}\kappa_{t}^{\rm c,g}) - E_{x,t}^{\rm c,g} \, e^{B_x^{\rm g}K_{t}^{\rm g}+\alpha_x^{\rm c,g}+\beta_x^{\rm c,g}\kappa_{t}^{\rm c,g} } \right)  {\ + \ \tilde{C}}.
\end{equation}
We impose the parameter restrictions $\|B^{\rm g}\|=\|\beta^{\rm c,g}\|=1$ {(with $\|\cdot\|$ the Euclidian norm)} and $\sum_{t\in{\cal T}} K_t^{\rm g} = \sum_{t\in{\cal T}} \kappa_t^{\rm c,g}=0$ when we maximize \eqref{eq:3a}-\eqref{eq:3} for all countries c and genders g to ensure that no identification issues can arise.

\paragraph{Estimated parameters for {the} Li-Lee baseline model.}

The estimated parameters are shown in Figure~\ref{fig:LiLee}. The $B_x^{\rm g}$ parameters show that the highest improvements in mortality in the observed period occurred at ages below 20 years. For higher ages the sensitivity to the $K_t^{\rm g}$ parameter is relatively stable, though ages 70-80 seem to benefit a bit more than average. The general trend in mortality across the countries, as represented by the slope of the $K_t^{\rm g}$ parameter, is stable, indicating that mortality (averaged over the countries) improved at a stable pace. The $\alpha_x^{\rm c,g}$ parameters exhibit the well-known structure: the lowest level of mortality at around 10 years, the accident hump for young adults, and mortality linearly increasing by age for higher ages. Differences between the countries are small except for ages close to the accident hump, and French female at higher ages seem to have structurally lower levels of mortality than the other countries considered.

The country-specific improvements, captured by the $\beta_x^{\rm c,g}$ and $\kappa_t^{\rm c,g}$ parameters, are less straightforward to interpret. The $\kappa_t^{\rm c,g}$ parameters indicate that countries may experience periods with higher and periods with lower improvement rates than the peer group. For example, the German male $\kappa_t$ parameter decreases from 1970 to 1985, then increases until 2010, and remains steady afterwards. Such periods of higher and lower improvement rates can alternate from one year to the other, as indicated by the abrupt change in slope in those period effects. The $\beta_x^{\rm c,g}$ parameters, which indicate how sensitive ages are towards changes in the $\kappa_t^{\rm c,g}$ parameter, are not consistently above or below zero. Therefore, we cannot say in general that for a specific dataset mortality increased less or more severely for all ages compared to the peer group.

\paragraph{Forecasting period effects in Li-Lee baseline model.}
We have used data until 2019 to calibrate the Li-Lee baseline model. To calibrate the new age-dependent factor and time series for the effect of the pandemic, we need to predict the baseline mortality rates for the years 2020 and 2021. In Section~\ref{sec:Forecasting} we investigate the impact of the pandemic on forecasts of cohort life expectancies, for which we need to forecast mortality far into the future. Hence, we need to specify a time series model for the calibrated period effects $K_t^{\rm g}$ and $\kappa_t^{\rm c,g}$.

The common approach to forecast the period effects in the peer group of countries is using a random walk with drift, see, for example, \cite{LiLee2005} and \cite{AG2020}. The assumption of a random walk with drift implies that the shared mortality trend continues in the future. For the country-specific period effects,  an autoregressive model of order one, including a constant, is often used. We aim to calibrate one large time series model for all period effects simultaneously. Preliminary analyses highlighted that for various country-specific period effects the mean reversion parameter is larger than one, indicating that the time series is not stationary. We therefore impose a random walk with drift for the country-specific period effects. For projection purposes, we neglect the drift term for the country-specific period effects to ensure mortality in the different countries does not diverge. This approach is justified by the fact that the estimated drift terms for most country-specific period effects are not statistically different from zero.

We define the assumed time series {for} the period effect of the peer group of countries and for the period effect of the country-specific deviations  as:
\begin{align*}
	K_t^{\rm g} &= K_{t-1}^{\rm g} + \theta^{\rm g} + \varepsilon_t^{\rm g}\\
	\kappa_t^{\rm c,g} &= \kappa_{t-1}^{\rm c,g} + \delta^{\rm c,g} + \xi_t^{\rm c,g}.
\end{align*}
We assume that the error terms $\varepsilon_t^{\rm g}$ and $\xi_t^{\rm c,g}$ for $ {\rm g}\in\{ {\rm m,f}\}$ and $ {\rm c}\in {\cal C}$ follow a multivariate normal distribution with mean vector $\textbf{0}_{12}$ and covariance matrix $\Sigma$. The parameters $\theta^{\rm g}$, $\delta^{\rm c,g}$ and all elements of the covariance matrix $\Sigma$ are estimated using maximum likelihood estimation techniques.

\subsection{Incorporation of the seasonal effect}\label{sec:IncorporateSeasonalEffect}

The specification for the weekly force of mortality in~\eqref{eq:1} contains a parameter for seasonal effect. We consider two approaches to incorporate a seasonal effect. If no seasonal effect is estimated in advance, which means $\ln\phi_{\x, w}^ {\rm c,g}$ is taken to be zero, it will be included in the time series $\mathfrak{K}_{t,w}^{\rm c,g}$. An alternative approach is to include a pre-determined estimator based on historical data which ensures that the product $\mathfrak{B}_\x^{\rm c ,g}\mathfrak{K}_{t,w}^{\rm c,g}$ only represents the observed deviation from the baseline predictions that are based on statistical information before the virus struck. We will designate the first approach with the term ``time series for seasonal effect plus COVID-19'' and the second approach with ``time series for COVID-19''. By comparing these two methods, we can investigate whether the two choices can lead to different conclusions about the impact of the pandemic.

\begin{figure}[!t]
\centerline{\includegraphics[width=\textwidth]{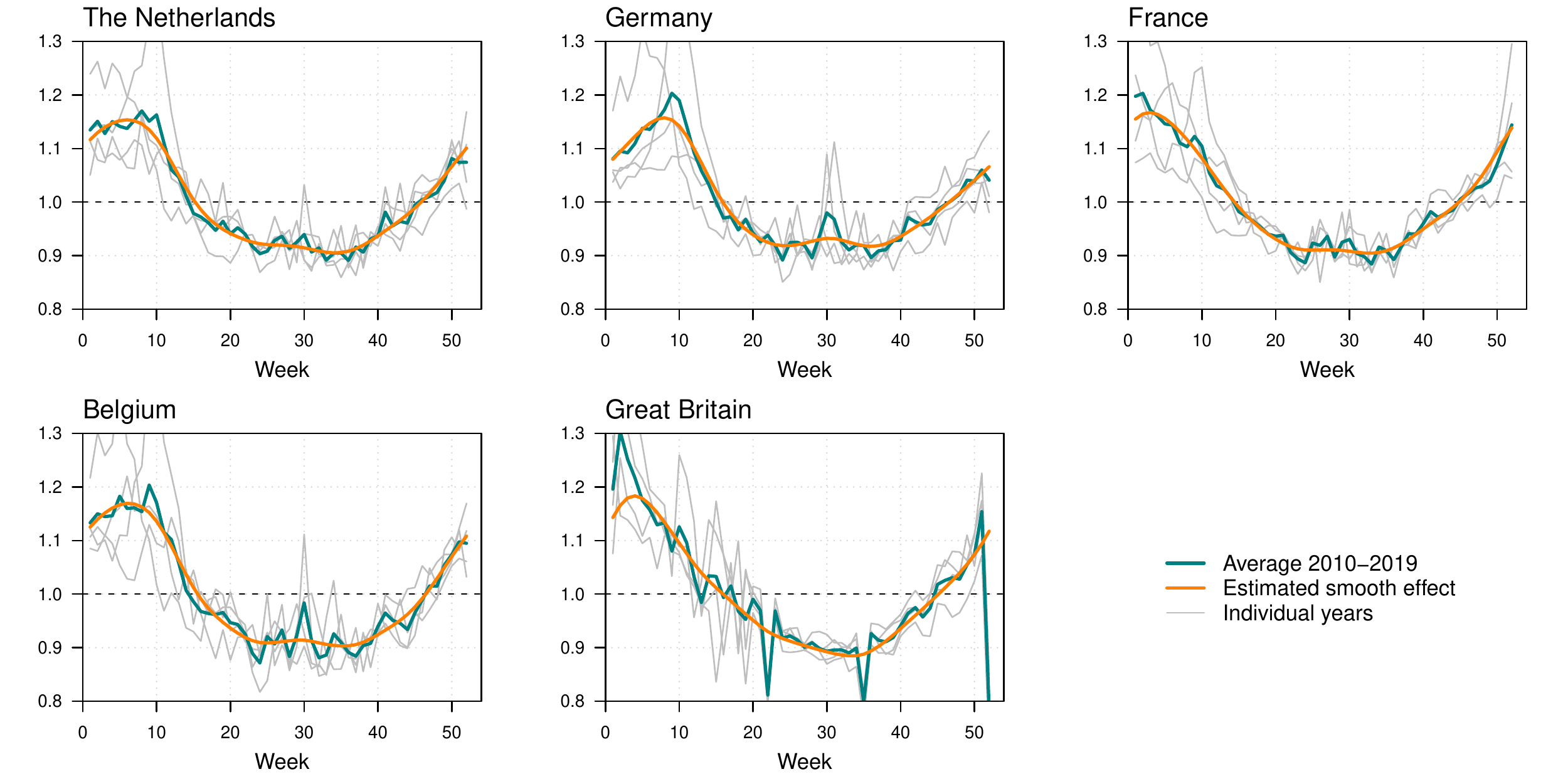}}
\caption{Observed fraction of annual mortality per week in the years 2010-2019 for the five countries considered, and estimated seasonal effect, aggregated across ages and sexes.}
\label{fig:season}
\end{figure}

To illustrate the second approach, Figure \ref{fig:season} shows how mortality spreads over different weeks of the year for the different countries. Values are aggregated over the two sexes and over all ages $x\in\cal X$. The gray lines for the years 2010 to 2019 show how mortality fluctuates over the weeks and we see a clear variation in time: for example, the severe flu wave around the tenth week of 2018 in the Netherlands is clearly visible in one of the gray lines. A value of $100\%$ in the figure corresponds to the situation in which mortality is uniformly distributed in a year. The green line is the average over all gray lines, and thus equals the observed average effect per week. Mortality during the year is not evenly distributed: as expected, more people die in the winter months than during the summer months. Using cyclic cubic splines, we have estimated a smooth effect based on the annual historical observations shown here, and the result is represented by the orange line. The values at the beginning and at the end of the year ensure a smooth transition from week $52$ to week $1$.


\subsection{Estimation of the impact of COVID-19}

Once the Li-Lee model for the baseline and the seasonal effects have been calibrated, we can estimate the remaining parameters $\mathfrak{B}_\x^{\rm c,g}$ and $\mathfrak{K}_{t,w}^{\rm c,g}$ that describe the effect of COVID-19. We make the assumption that during the pandemic the number of deaths for given exposures still follows a Poisson distribution which now includes the additional factors. This implies that for $t=2020$ and $t=2021$ we impose
$$
	D_{\x,t,w}^{\rm c,g}\sim {\rm Poisson}\left(\, E_{\x,t,w}^{\rm c,g} \, \mu_{\x,t}^{\rm c,g} \, \phi_{\x,w}^{\rm c,g} \exp({\mathfrak{B}_\x^{\rm c,g}\mathfrak{K}_{t,w}^{\rm c,g}})\,\right)
$$
(where $\mu_{\x,t}^{\rm c,g}$  and $\phi_{\x,w}^{\rm c,g}$ are set to {the earlier} calibrated values), and the required dataset for calibration is 
$$
	\{(E_{\x,t,w}^{\rm c,g},\, D_{\x,t,w}^{\rm c,g}),\ \x\in{\cal X}, \ {\rm g}\in\{ {\rm m,f}\},\  {\rm c}\in {\cal C},\ t\in {\cal T},\ w\in {\cal W}_t\}.
$$
We choose a subset ${\cal X}$ of ages and the collections ${\cal T}=\{2020,2021\}$ and ${\cal W}_{2020}=\{1,..,53\}$ and ${\cal W}_{2021}=\{1,..,52\}$ for weeks\footnote{The year 2020 counted among the used calendar convention NEN 2772/ISO 8601 an (incomplete) 53$^{rd}$ week and 2021 an (incomplete) 0$^{th}$ week; a correction has been made in the datasets by merging the two.} after January 1st, 2020.
The parameters that describe the impact of the pandemic thus follow by determining for $ {\rm g}\in\{ {\rm m,f}\}$ and  $ {\rm c}\in {\cal C}$:
\begin{eqnarray}\label{eq:maxll}
( \widehat{\mathfrak{B}}_{\x}^{\rm c,g}, \widehat{\mathfrak{K}}_{t,w}^{\rm c,g})
&=& \arg\min_{(\mathfrak{B}_{\x}, \mathfrak{K}_{t,w})}\,
\sum_{\x\in\cal X}
\sum_{t\in {\cal T}}
 \sum_{w\in{\cal W}_t}
\left(D_{\x,t,w}^{\rm c,g} \mathfrak{B}_{\x}\mathfrak{K}_{t,w} - D^{\rm pred,g,c}_{\x,t,w}\exp(\mathfrak{B}_{\x}\mathfrak{K}_{t,w}) \right),
\end{eqnarray}
{where} we define
\begin{equation}\label{eq:pred}
	D^{\rm pred,c,g}_{\x,t,w}= E^{\rm c,g}_{\x,t,w}  \mu_{\x,t}^{\rm c,g} \, \phi_{\x,w}^{\rm c,g}
\end{equation}
for the expected number of deaths in a certain week based on given exposures during the week and the baseline mortality calibration, while possibly taking into account a seasonal effect. 


\paragraph{Estimation of weekly deaths and exposures by age.}

The weekly exposures  during the pandemic in \eqref{eq:pred}, i.e., the values of $E^{\rm c,g}_{\x,t,w}$, cannot be based on observed data so we must generate estimates based on the population data that we have at our disposal.\footnote{To generate estimates of weekly exposures, we will make the assumption that each year consists of 52 weeks, the month February consists of 28 days, and the month December consists of 30 days. This results in a year of 364 days, which equals exactly 52 weeks. This approach has some clear drawbacks. For example, week 1 does not start on 1 January in each year, and December obviously has 31 days. However, this approach does take into account the development in the population estimates through the year. In case a year has 53 weeks, we assume that the exposure in week 53 is the same as it is in week 52 of that year. The exposure for a certain week is estimated as the average population during that week multiplied by $7/365$, so in the sequel we can focus on the determination of the population numbers per day during $2020$ and $2021$.} 
Various population statistics are available in the Eurostat database. Unfortunately, only yearly population estimates are given, and the most recent population estimate at the time that this paper was written refers to the measurements on 1 January 2020 for $c\in\{\rm NLD, BEL, DEU, FRA\}$ and 1 January 2019 for $c=\rm GBR$.
 
{We determine the population on January 1 of year $t+1$ as} $P_{x,t+1}=P_{x-1,t} - C_{x,t}$, where $C_{x,t}$ denotes the number of people that died in year $t$ that would have had age $x$ at 31 December of that year; these data can be obtained from the Eurostat database\footnote{We used a backtest over the years $2015$ up to, and including, $2019$ to investigate whether this is a good method to predict the change in population figures for an individual age over the course of a year, and found satisfactory results. Including migration information improves the accuracy of this method, but weekly migration information is not available and is therefore not used when estimating the weekly population sizes.}. Some adjustments are necessary to estimate weekly exposures. We combine the population sizes from Eurostat with available weekly mortality observations per age-group, $D_{\x,t,w}^{\rm c,g}$, from the {\sl Short Term Mortality Fluctuations} dataset, which works with age-groups of five years, {starting from} $\{0,1,2,3,4\}$ until $\{85,86,87,88,89\}$, and an open-ended final age-group $\{90,91,92,...\}$. 
We transform the values $D_{\x,t,w}^{\rm c,g}$ for age-groups $\x$ to values $D_{x,t,w}^{\rm c,g}$ for individual ages $x$ in $t=2020$ and $t=2021$ by assuming that the proportional distribution of deaths over ages in an age-group in a certain week $w$ equals the historical average of that distribution for age-group over the years $2015$ up to (and including) $2019$:
\begin{equation}
	D_{x,t,w}^{\rm c,g} = D_{\x,t,w}^{\rm c,g} \cdot \frac{\sum_{t=2015}^{2019} D_{x,t}^{\rm c,g}}{\sum_{x\in\x}\sum_{t=2015}^{2019} D_{x,t}^{\rm c,g}}. \label{eq:deaths_age_group}
\end{equation}
Define $C_{x,t,w}^{\rm c,g}$ as the number of people that died during week $w$ in year $t$ that would have had age $x$ at 31 December of year $t$. We construct $C_{x,t,w}^{\rm c,g}$ using the approximation\footnote{This approximation is based on the assumption that at all times mortality is uniformly spread over all people in a Lexis-parallelogram who have the same age at the end of the year, and uniformly spread during the year over all people with a common (rounded) age.}: 
$$
	C_{x,t,w}^{\rm c,g} = \left(1-\frac{w}{w_t}\right) \cdot D_{x-1,t,w}^{\rm c,g} + \frac{w}{w_t} \cdot D_{x,t,w}^{\rm c,g},
$$
where $w_t$ represents the number of weeks in year $t$, so $w_{2019}=52$, $w_{2020}=53$, and $w_{2021}=52$.
We could estimate the population size at the first day of week $w+1$ using 
$$ 
	P_{x,t,w+1}^{\rm c,g} = P_{x,t,w}^{\rm c,g} - \left(\tfrac{w}{w_t}\cdot C_{x+1,t,w}^{\rm c,g} + (1-\tfrac{w}{w_t})\cdot C_{x,t,w}^{\rm c,g}\right).
$$
However, we then ignore the fact that people may have their birthday during a week, and we therefore replace the above formula by
\begin{equation}
	P_{x,t,w+1}^{\rm c,g} =  \left(1-\frac{w}{w_t}\right) \left( P_{x,t,1}^{\rm c,g} -\sum_{i\le w} C_{x+1,t,i}^{\rm c,g} \right) +
						\frac{w}{w_t}\left( P_{x-1,t,1}^{\rm c,g} -\sum_{i\le w} C_{x,t,i}^{\rm c,g} \right), \label{eq:ProjP}
\end{equation}
which accounts for this effect under the assumption that births are distributed uniformly during the year.  We initialize this procedure using the last known value of $P_{x,t}$ and we apply this formula for $w=0,...,w_{t}$, such that $P_{x,t,1}=P_{x,t}$ and $P_{x,t,w_t+1}=P_{x,t+1,1}\approx P_{x,t+1}$.

Once we have a value for the population at the beginning and the end of the week, we can take the average to find the average population during the week, and we then have the required exposure $E^{\rm c,g}_{x,t,w}$ for that week after we have multiplied the result by $\tfrac{7}{365}$.

For the Netherlands, we have more granular data at our disposal. Population sizes are available for the first day of each month until 1 January 2022. For the other days in the month, we determine the population sizes through linear interpolation. We do not need to project population sizes as in~\eqref{eq:ProjP}, since linear interpolation over monthly estimates is more accurate then projecting monthly estimates over a two-year period. The weekly exposures are then determined by multiplying the average population during the week by $\tfrac{7}{365}$. The mortality observations $D_{x,t,w}^{\rm c,g}$ are available for individual ages and can be used directly.

\section{Empirical Results}\label{sec:Results}

In this section, we present the empirical results when the model introduced in the previous section is applied to actual data. First, we analyze the impact of COVID-19 on mortality observations without using exposure information. This approach can be applied relatively easily, since during pandemics mortality observations may quickly become available, whereas exposure observations are often estimated and published on an annual basis only. We calibrate the COVID-19 model to Dutch mortality data for which we have mortality data available for individual ages. We use this granular dataset to investigate different ways to include a seasonal effect, analyze which ages are most affected by COVID-19, and examine the importance of having granular data. Finally, we compare the impact of COVID-19 on the level of mortality over time for the collection of five countries: NLD, DEU, FRA, BEL, and GBR.

\subsection{Analysis based on Dutch death counts only}

In this section we show, as comparison, the impact of COVID-19 using only the number of deaths during the weeks of 2020 and 2021. This avoids the use of exposures which may be hard to estimate. We follow \cite{Oeppen2008} in which the Compositional Data (CoDa) analysis is introduced, referring to \cite{Aitchison}, as an alternative to the Lee-Carter way of modeling mortality. In \cite{Oeppen2008}, the CoDa analysis is presented in a number of steps which make use of so-called CoDa operators. These operators are summarized in an appendix of \cite{BergeronEtAl2017}. We follow this approach and apply it to Dutch mortality data for the years 2010 to 2021 and ages 0 to 98. The CoDa analysis cannot deal with zero observations. Since the weekly death counts by age contain many zero observations, we increase all death counts by 1. 
This adjustment has negligible impact on the {shape of the} distribution of the deaths over the ages and over the weeks.


\begin{figure}[h]
\centerline{\includegraphics[width=\textwidth]{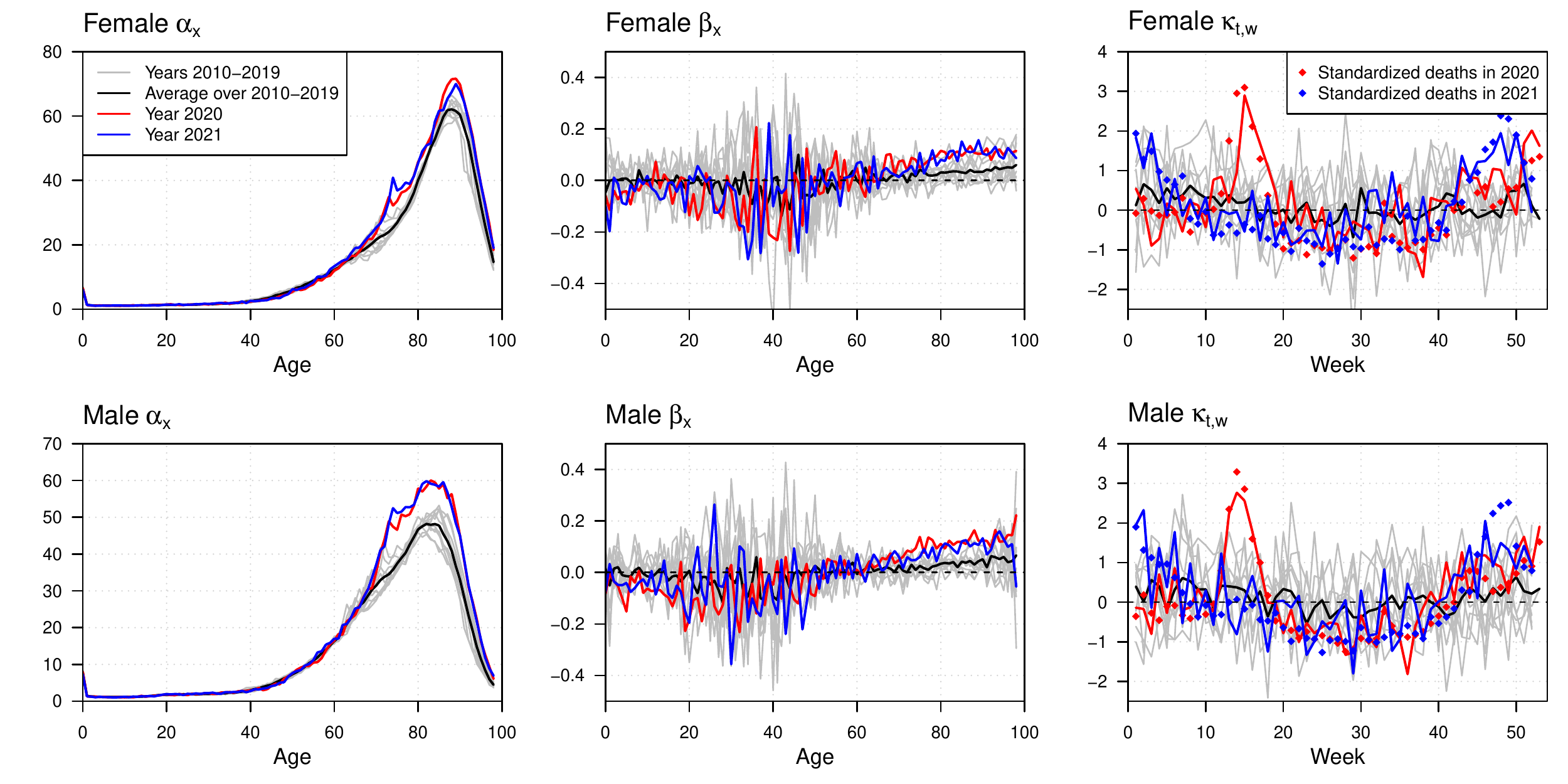}}
\caption{Results from CoDa analysis using Dutch weekly number of deaths. The gray lines show the estimated CoDa parameters for the years 2010 to 2019 and the black line represents the average over these estimates. The estimated CoDa parameters for the years 2020 and 2021 are shown in red and blue respectively. For the years 2020 and 2021, the graphs on the right also show the standardized weekly death counts (normalized using the mean and variance in that year).}
\label{fig:coda}
\end{figure}

Figure \ref{fig:coda} shows the CoDa-parameter estimates. In the CoDa analysis applied to mortality data, the $\alpha_x$ parameters show the (average) age distribution of the death counts.\footnote{This age distribution shows the absolute number of deaths per age, aggregating to the total number of deaths per week.} These age distributions in 2020 and 2021 are close to the average over the years 2010 to 2019 for lower ages. For higher ages, however, the age distributions in 2020 and 2021 are higher than the average over the years 2010-2019 and in particular also the age distribution in 2019. This confirms that COVID-19 mainly increased the number of deaths of the elderly. The age distribution of 2021 is close to that of 2020, although there seems to be a slight shift to younger ages. In particular, there seems to be a small increase for the ages between 50 and 70.

The $\kappa_{t,w}$ process in the CoDa analysis shows shifts in the age distribution over the weeks. By construction, both the $\kappa_{w}$ process (summed over time) and the $\beta_x$ parameters (summed over ages) add up to zero. A positive value of $\kappa_{w}$ implies a shift in the age distribution of number of deaths from the ages with a negative $\beta_x$ to the ages with a positive $\beta_x$. The $\kappa_{w}$ process shows a clear peak between weeks 10 and 20 of the year 2020, the first wave of the COVID-19 pandemic in the Netherlands. It also shows increasing positive values near the end of 2020, followed by positive values at the start of 2021, the second wave of COVID-19. Finally, near the end of 2021 the $\kappa_{w}$ process shows the third wave of COVID-19. {For comparison we also show the standardized weekly death counts, i.e., the death counts normalized using the mean and variance in that year.}

These peaks imply a shift in the age distribution from the young to the old, since the $\beta_x$s of the older people (above age 60) are positive, while the $\beta_x$ values of the younger ages are mostly negative (although there are some exceptions as far as the younger ages are concerned). However, the $\beta_x$s for the older males are substantially higher in 2020 than in 2021. {The} shifts in the age distribution (corresponding to an increase in the $\kappa_w$ process) of the males in 2021 {are} far less dramatic than the shifts in 2020. For females, the $\beta_x$ coefficients of 2021 are close to those in 2020 for ages above 80, but for ages between 60 and 80 the $\beta_x$ values in 2021 are lower than those in 2020. Thus, the shifts in the age distribution for females in 2021 is particularly in the direction of the very old, above age 80. 


\subsection{Calibration results based on a Dutch dataset with high granularity}

\noindent In this section we continue our analysis using the Dutch mortality data as obtained from Statistics Netherlands.
First, we investigate the two different approaches for incorporating the seasonal effect, then we compare the results when using different age ranges for calibration, and finally we analyze the importance of using granular data.

\paragraph{Incorporation of seasonal effect.}

\begin{figure}[!h]
\centerline{\includegraphics[width=\textwidth, trim = 0cm 0.0cm 0cm 0cm]{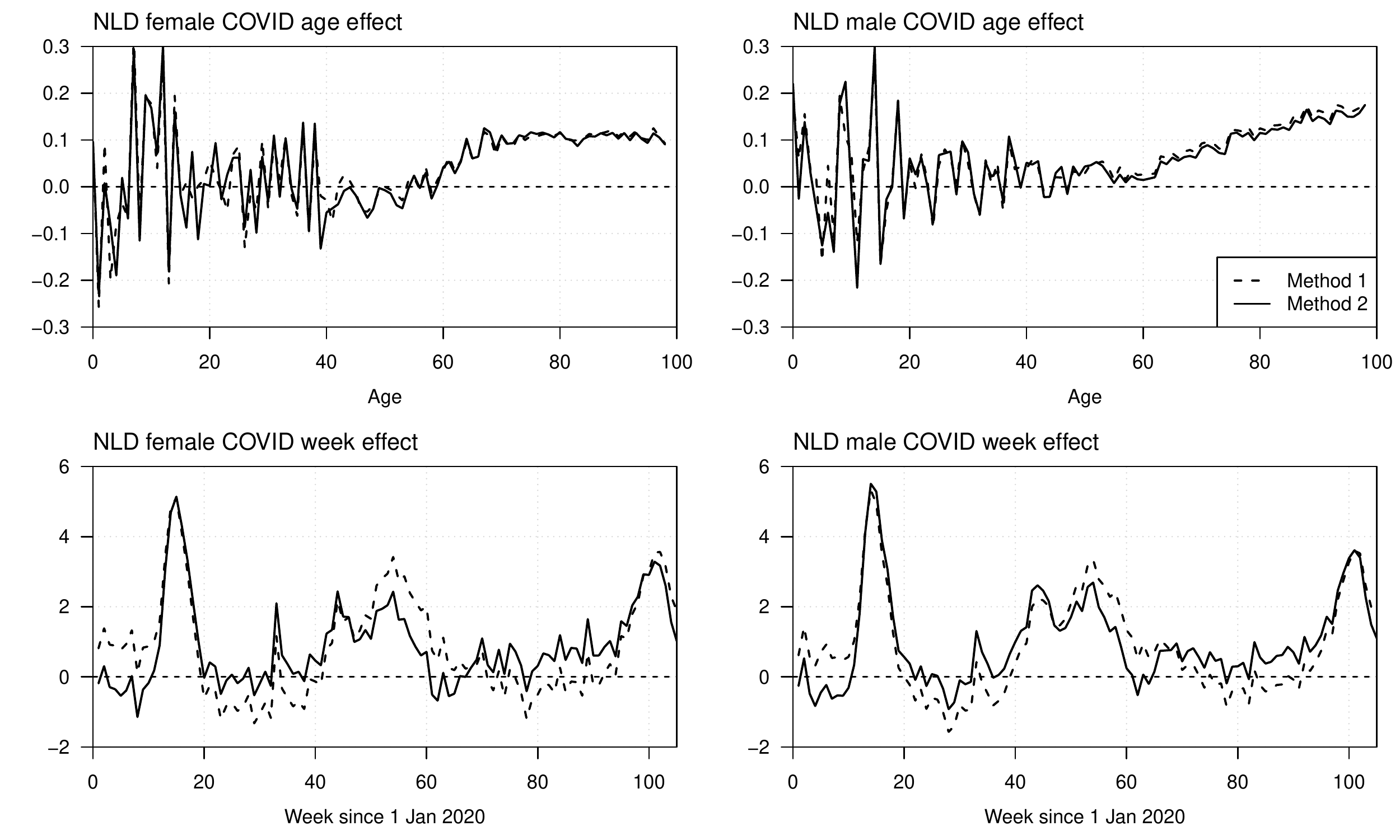}}
\caption{COVID-19 parameters estimated with and without predetermined seasonal effect (solid respectively dashed line) for females and males using Dutch data.}
\label{fig:NLD_parameters_seasonal}
\end{figure}

In Section~\ref{sec:IncorporateSeasonalEffect} we described two approaches for including the seasonal effect:
\begin{description}
	\item[Method 1] \textit{Time series for seasonal effect plus COVID-19}: the seasonal effect $\phi_{\chi,t}^{\rm c,g}$ in~\eqref{eq:1} is set equal to 1, which results in any seasonal effect being captured by the COVID-19 term $\mathfrak{K}_{t,w}^{\rm c,g}$.
	\item[Method 2] \textit{Time series for COVID-19}: the seasonal effect $\phi_{\chi,t}^{\rm c,g}$ is set equal to the estimated smooth seasonal effect as illustrated in Figure~\ref{fig:season}, which results in the COVID-19 term $\mathfrak{K}_{t,w}^{\rm c,g}$ reflecting the impact of COVID-19 corrected for this seasonal effect.
\end{description}
Figure~\ref{fig:NLD_parameters_seasonal} shows the calibrated parameters. We observe that the differences in the COVID-19 age effects, as shown in the top row, are hardly visible. The COVID-19 age effect is erratic and close to zero for both females and males up to age 60. This indicates that for those ages there was hardly any impact on the level of mortality due to COVID-19 in the years 2020 and 2021. For higher ages, the age effect for females increases from age 60 to 70 and remains constant for ages 70 to 98, and for males the age effect increases steadily from age 60 to 98.

The COVID-19 week effects (bottom row) show substantial differences between the two methods. In the Netherlands, the corona virus was first identified in February 2020, which means that in the first four weeks of 2020 we would expect the COVID-19 week effect to be close to zero. The week effect represented by the dashed lines clearly starts above 0 for males and females. Further, in the summer of 2020 (around week 26) the dashed lines for the week effect are below zero, indicating mortality levels were below expectation if seasonality was not taken into account. The COVID-19 week effect of Method 2 (the solid line) starts close to zero and remains above zero for nearly all weeks included in the dataset; Method 2 therefore seems to capture only the impact of COVID-19 on observed mortality. 

\begin{figure}[!h]
\centerline{\includegraphics[width=\textwidth, trim = 0cm 0.0cm 0cm 0cm]{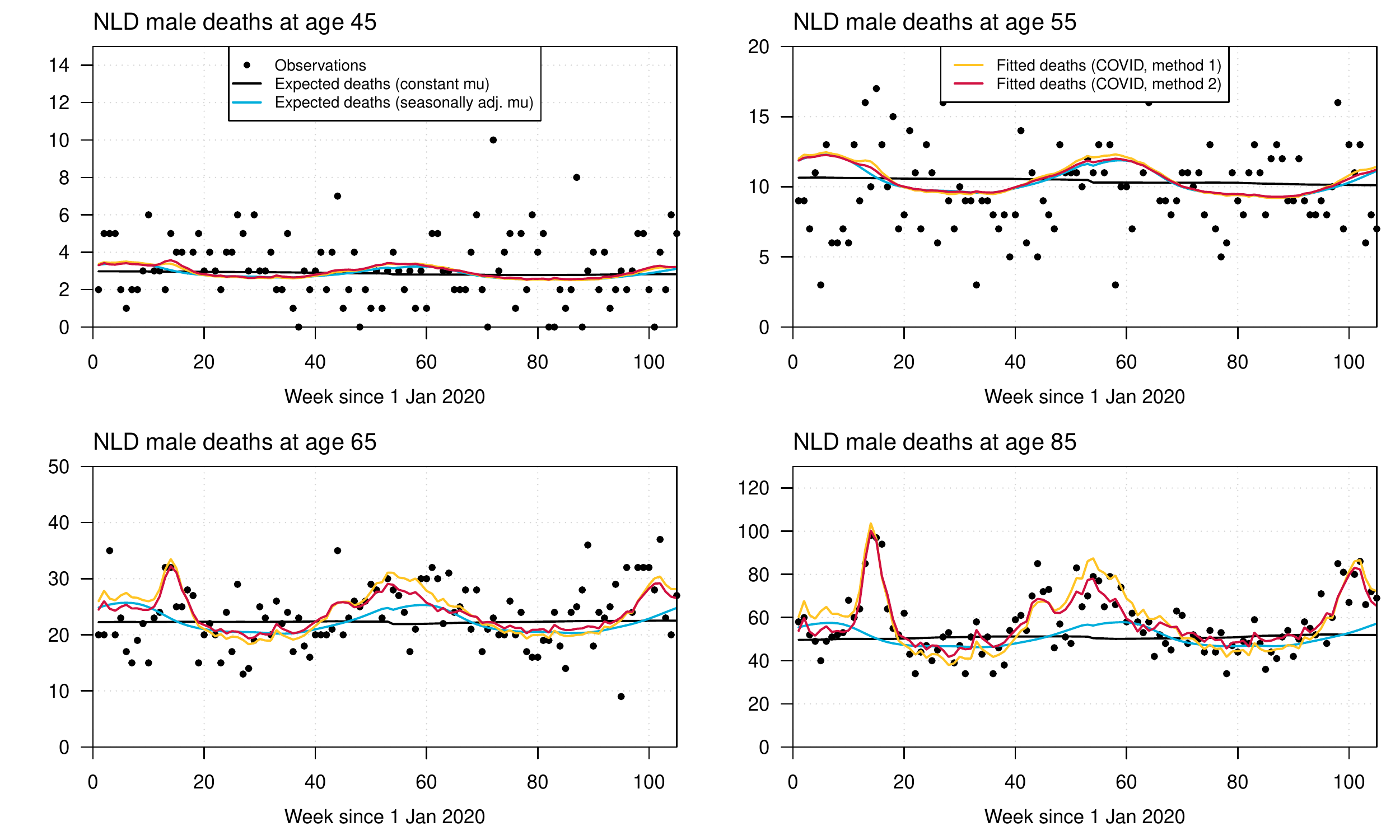}}
\caption{The black dots represent observed weekly deaths for Dutch males at ages 45, 55, 65 and 85 during the years 2020 and 2021. The black line shows the expected deaths assuming a constant force of mortality, and the blue line represents expected death counts taking into account the historical seasonal effect. The yellow and red lines show the fitted deaths using Method 1 respectively Method 2.}
\label{fig:NLD_fitted_mortality}
\end{figure}

The observed weekly mortality death counts as shown in Figure~\ref{fig:NLD_fitted_mortality} are very volatile for the ages 45 and 55 and do not exhibit the peaks as observed in the COVID-19 week effects in Figure~\ref{fig:NLD_parameters_seasonal}. The fitted death counts are close to the seasonally adjusted death counts, and the impact of COVID-19 on mortality at these ages is therefore negligible. For ages 65 and 85, the seasonally adjusted expected deaths clearly exhibit a seasonal pattern, but this pattern is not sufficient to capture the peaks that coincide with the COVID-19 waves. The fitted deaths using Method 1 and Method 2 follow the wave pattern in the observed deaths more closely.

We would like to be able to distinguish between effects induced by a typical seasonal effect and effects due to a pandemic. Method 2, in which we correct for historically observed seasonal effects, allows us to assess the impact of only the pandemic on the level of mortality. Further, Method 2 is more flexible than Method 1, since the seasonal effect is not enforced to affect mortality the same way as the pandemic. Therefore, in the remainder of the analyses we will only use Method 2 when estimating the impact of COVID-19 on the level of mortality.

\paragraph{Selection of ages.}

The COVID-19 age effects for ages below 40 in Figure~\ref{fig:NLD_parameters_seasonal} are volatile and seem to be centered around zero. In Appendix \ref{sec:selection_of_ages} we show estimates using ages 0-98 and ages 40-98. From the parameter estimates in Figure~\ref{fig:NLD_parameters_age_selection}, we conclude that the COVID-19 week parameters and age parameters for higher ages are hardly affected when including information from younger ages. For interpretation of parameters and reliability of estimates and projections, it is desirable that only structural effects are analyzed; noise due to low exposures should preferably be excluded. Since the age effects at the younger ages do not seem to capture a systematic COVID-19 effect, we exclude data for ages below 40 from calibration.

\paragraph{Importance of granular data.}

For the Netherlands we have weekly mortality data available for both genders and individual ages. The \textit{Short Term Mortality Fluctuations database} (STMF) contains weekly mortality data for many countries, but only for specific age groups. For some countries the age groups are small (age groups of five years), whereas for other countries there are only five age groups.

In this section, we use the dataset from Statistics Netherlands to investigate the importance of having granular data in analyzing the impact of COVID-19 on mortality. We consider three levels of granularity:
\begin{itemize} 
	\item \textbf{Level 1}, individual ages with $\x\in\{0,1,2,...,99\}$; this is the most granular type of data;
	\item \textbf{Level 2}, age groups of five years with {$\x\in\{\{0,...,4\},\{5,...,9\},...,\{90,...,94\},\{95,...\}\}$};
	\item \textbf{Level 3}, five age groups with {$\x\in\{\{0,...,14\},\{15,...,64\},\{65,...,74\},\{75...84\},\{85,...\}\}$}; this is the least granular type of data.
\end{itemize}
We use the original dataset with observations by individual ages (Level 1) to artificially construct the datasets based on Level 2 and Level 3 granularity.

\begin{figure}[!h]
\centerline{\includegraphics[width=\textwidth, trim = 0cm 0.0cm 0cm 0cm]{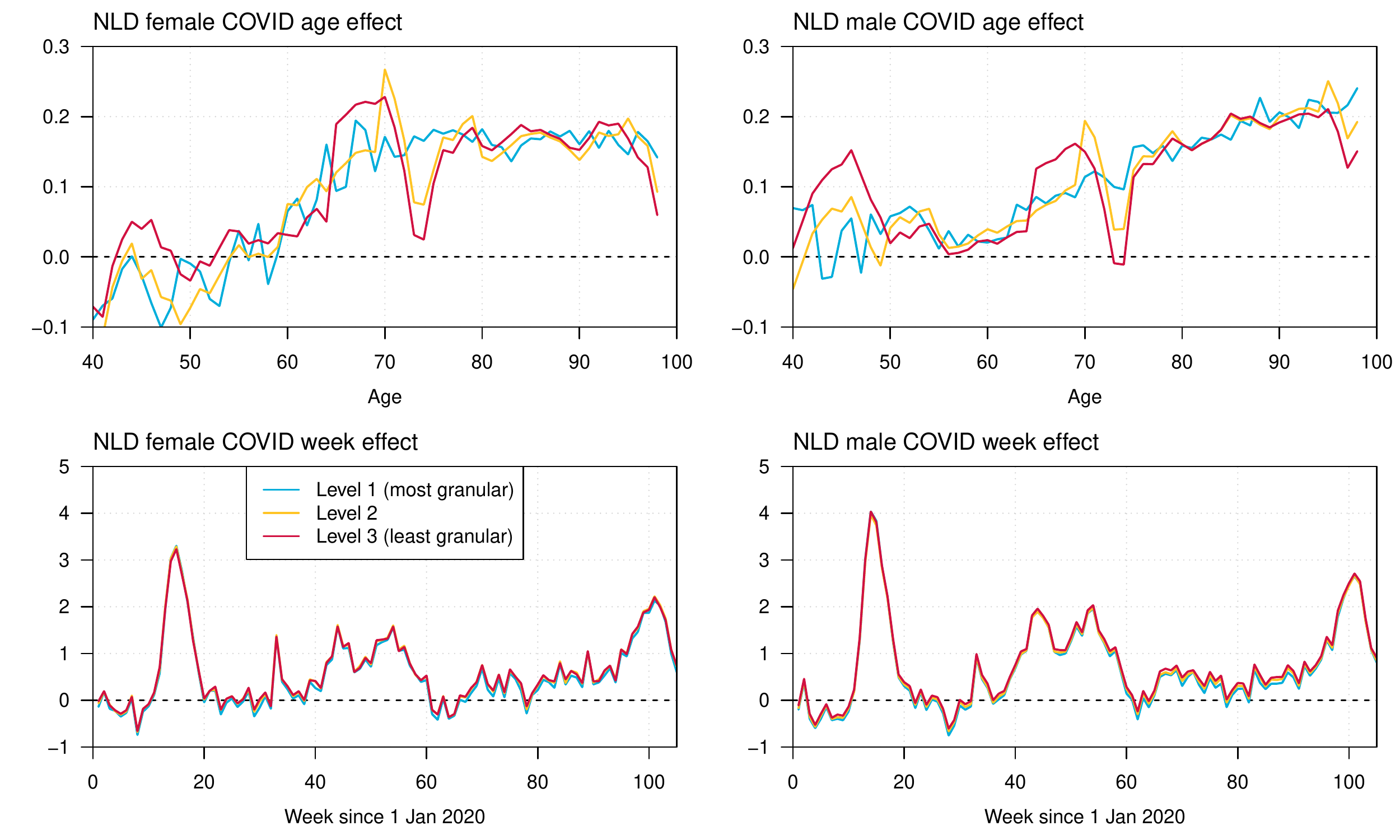}}
\caption{COVID-19 parameter estimates using datasets with different levels of granularity.}
\label{fig:NLD_importance_granularity}
\end{figure}

Figure~\ref{fig:NLD_importance_granularity} shows the parameter estimates when calibrating the COVID-19 model to the three datasets. The top row shows the estimated COVID-19 age effects, and we observe clear differences between the estimates from the datasets with different levels of granularity. The age effects are volatile for all levels of granularity, but the Level 2 and Level 3 estimates show remarkable peaks and dips at the ages 65-75. In the year 2020, these ages correspond with the years of birth 1945-1955 {(with a clear peak in 1946)}, which is the so-called baby boom generation after the second World War. The baby boom generation appears as a cohort effect in the exposures, which in turn results in cohort effects in the observed deaths. For most age groups (either Level 2 or Level 3), the distribution of deaths over the ages is relatively stable over time, except for this baby boom generation. This cohort effect makes the allocation of deaths to individual ages increasingly inaccurate if larger age groups and more historical years are used when applying Equation~\eqref{eq:deaths_age_group}.

In contrast, the COVID-19 week effects estimated using the datasets of different granularity look remarkably similar. This is as expected since the observed deaths are not relocated over different weeks, only over different ages. The impact of COVID-19 on total mortality will therefore be similar, which results in stable estimates of the COVID-19 week effect, regardless of the level of granularity of the dataset.

From this analysis we conclude that datasets with deaths by age group can be used to {obtain a first impression of} the impact of a pandemic over time. However, if the impact {on individual ages must be assessed, one needs to use death counts for} individual ages.

\subsection{Calibration results based on the STMF dataset}

In this section, we calibrate the COVID-19 model to the countries $c$ in the set given by $\mathcal{C}=\{\rm NLD, DEU, FRA, BEL, GBR\}$ using mortality data as obtained from STMF. The deaths by age group are allocated to individual ages using Equation~\eqref{eq:deaths_age_group}, and we include data for the ages 40 to 95. We investigate to what extent similarities and differences between countries can be observed.

\begin{figure}[!h]
\centerline{\includegraphics[width=\textwidth, trim = 0cm 0.0cm 0cm 0cm]{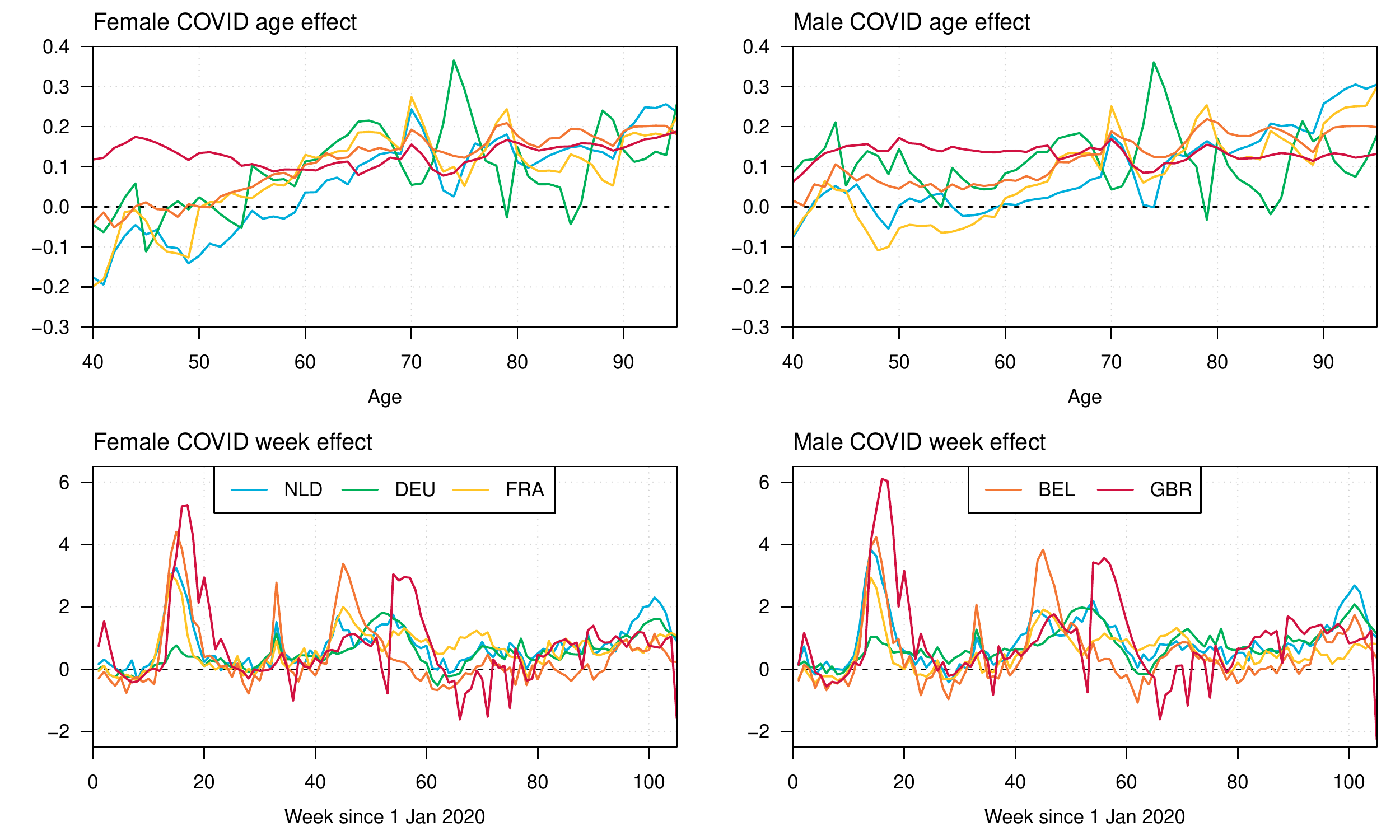}}
\caption{Estimated COVID-19 parameters for various countries using ages 40 to 95.}
\label{fig:FiveCountriesSelAges}
\end{figure}

Figure~\ref{fig:FiveCountriesSelAges} shows the estimated COVID-19 parameters.\footnote{Results obtained using ages 0-98 are available in Appendix~\ref{sec:STMF_all_ages}.} The COVID-19 age effect is close to zero at lower ages for most countries. The exception to this is Great Britain, where the COVID-19 age effect is substantial for all ages included. The volatile behavior between ages 65 and 80 is probably the result of cohort effects in the death counts, as described in the previous section. 

We observe that the different countries exhibit similar COVID-19 week effects, though there are a few notable differences. Germany experienced a minor first COVID-19 wave compared to the other four countries, whereas later COVID-19 waves were similar to those in other countries. Around week 33, Belgium, the Netherlands and Germany experienced a temporary peak, which may be the result of temporarily relaxing COVID-related restrictions.

Between week 40 and 60 (winter season 2020-2021) the second COVID-19 wave hit Western Europe, but we observe substantial differences between countries. From week 50 onward, COVID-19 vaccines became available in various countries, though the availability and timing of the vaccine shots for people of different ages differed between countries. Belgium had a high peak at around week 45 after which excess mortality soon disappeared. The Netherlands, Germany and France all experienced lower peaks, but in these countries it also took a few more weeks before excess mortality had vanished. This decrease in excess mortality may have been the result of vaccines being applied to parts of the population but also due to new COVID-19 restrictions. Finally, Great Britain experienced a high peak which was similar to the one in Belgium, but a few weeks later, and measures were taken resulting in excess mortality decreasing rapidly. Great Britain was one of the countries in which vaccines were provided to the population earlier and faster.

It will always remain challenging to assess the effectiveness of policy decisions during a pandemic. But using the best possible data and using improved estimation methods can help to improve such assessments.

\section{Forecasting mortality adjusted for COVID-19}\label{sec:Forecasting}

In the previous section we have analyzed the impact of COVID-19 on the level of mortality in the years 2020 and 2021. For insurance companies and pension funds it is particularly important to assess the impact of COVID-19 on mortality rates. At the time this paper was written, it was too early to predict whether long term mortality improvement rates should be adjusted upward, downward, or not at all. In this section, we therefore introduce a general framework in which the impact of COVID-19 on the level of mortality in the years 2020 and 2021 can be used to generate scenarios for future mortality.

\paragraph{Transforming weekly effects to annual effects.}

The impact of COVID-19 was calibrated using weekly mortality observations. From these calibrations, we obtained weekly parameter estimates $\mathfrak{B}_x$ and $\mathfrak{K}_{t,w}$. To generate scenarios for future mortality, we first need to transform these week effects into annual effects $\mathfrak{V}_x$ and $\mathfrak{X}_{t}$. We temporarily impose that $\sum_{x \in \cal X} \mathfrak{V}_x = 1$ such that analytical results can be used for the transformation. 
Afterwards, we again apply the restriction $\| \mathfrak{V} \|=1$. To transform week effects into annual effects, we make the two one-year survival probabilities for the years 2020 and 2021 equal to the product of the weekly survival probabilities in those years:
\begin{equation}\label{eq:week_annual}
\exp\left(- \mu_{x,t} \cdot \exp[\mathfrak{V}_x \mathfrak{X}_t] \right) =
 \prod_{w=1}^{w_t} \exp\left(
- \tfrac{1}{w_t} \cdot \mu_{x,t} \cdot \phi_{x,w} \cdot \exp\left[\mathfrak{B}_x  \mathfrak{K}_{t,w}\right]\right),
\end{equation}
for $t=2020$ and $t=2021$, and for all $x\in \cal X$. After taking the natural logarithm on both sides, dividing by $\mu_{x,t}$, taking the natural logarithm again, and summing over all ages $x\in \cal X$, we find
$$
\mathfrak{X}_{t} = \sum_{x\in\cal X} \ln \left\{ \tfrac{1}{w_t} \sum_{w=1}^{w_t} 
 \phi_{x,w} \cdot \exp[\mathfrak{B}_x  \mathfrak{K}_{t,w}]\right\}.\nonumber
$$
Next, we determine $\mathfrak{V}_x$ by making survival over both the years 2020 and 2021 equal to surviving over all weeks in those years (which follows directly from~\eqref{eq:week_annual}):
$$
\prod_{t=2020}^{2021}
\exp\left(- \mu_{x,t} \cdot \exp[\mathfrak{V}_x \mathfrak{X}_{t}] \right) =
\prod_{t=2020}^{2021} \prod_{w=1}^{w_t} \exp(
- \tfrac{1}{w_t} \cdot \mu_{x,t} \cdot \phi_{x,w} \cdot \exp[\mathfrak{B}_x  \mathfrak{K}_{t,w}]).\nonumber
$$
Rewriting this equation results in 
$$
\sum_{t=2020}^{2021}  \mu_{x,t} \sum_{w=1}^{w_t} \tfrac{1}{w_t} \left(\exp[\mathfrak{V}_x \mathfrak{X}_{t}] -
  \phi_{x,w} \cdot \exp[\mathfrak{B}_x \mathfrak{K}_{t,w}]\right) = 0,\nonumber
$$
and this non-linear equation in $\mathfrak{V}_x$ can be solved numerically for each age $x \in \cal X$ separately. Finally, we again renormalize the parameters $\mathfrak{V}_x$ and $\mathfrak{X}_t$ such that $\| \mathfrak{V} \|=1$.

\begin{figure}[!h]
\centerline{\includegraphics[width=\textwidth, trim = 0cm 0.0cm 0cm 0cm]{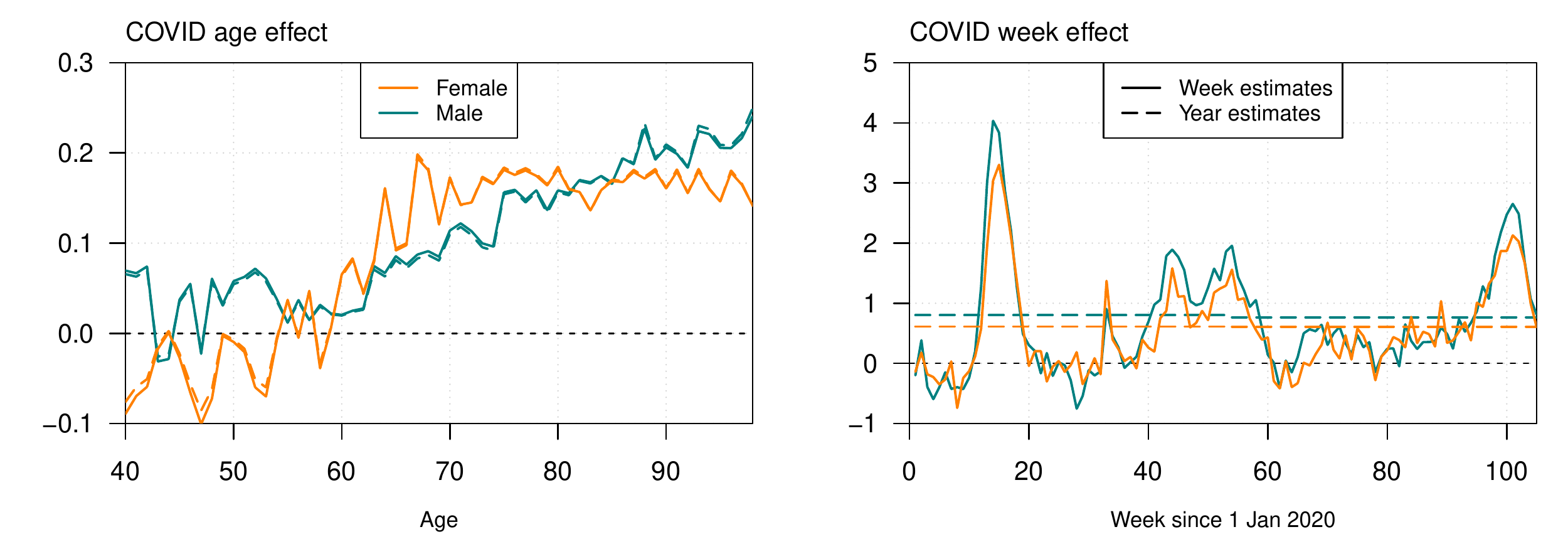}}
\caption{COVID-19 parameters from calibration on weekly data (solid line) and annualized COVID-19 parameters (dashed line).}
\label{fig:YearTransform}
\end{figure}

Figure~\ref{fig:YearTransform} shows the weekly and annual effects calibrated using the Dutch mortality data from Statistics Netherlands. The weekly and annual COVID-19 age effects are nearly identical, and the COVID-19 year effects are close to the averages of the week effects in that year.

\paragraph{Future scenarios for the COVID-19 period effect.}

At the end of the observed period, the week effect as shown in Figure~\ref{fig:YearTransform} is  {not far from} zero. For prediction purposes, one might be tempted to choose $\mathfrak{X}_{t}$ equal to 0 for $t>2021$. However, the week effect also shows that after periods of (almost) no excess mortality, new periods of excess mortality may occur.
 {We therefore} investigate how mortality rates  {could} develop under varying assumptions for the future impact of COVID-19 on mortality.

\begin{figure}[!h]
\centerline{\includegraphics[width=\textwidth, trim = 0cm 0.0cm 0cm 0cm]{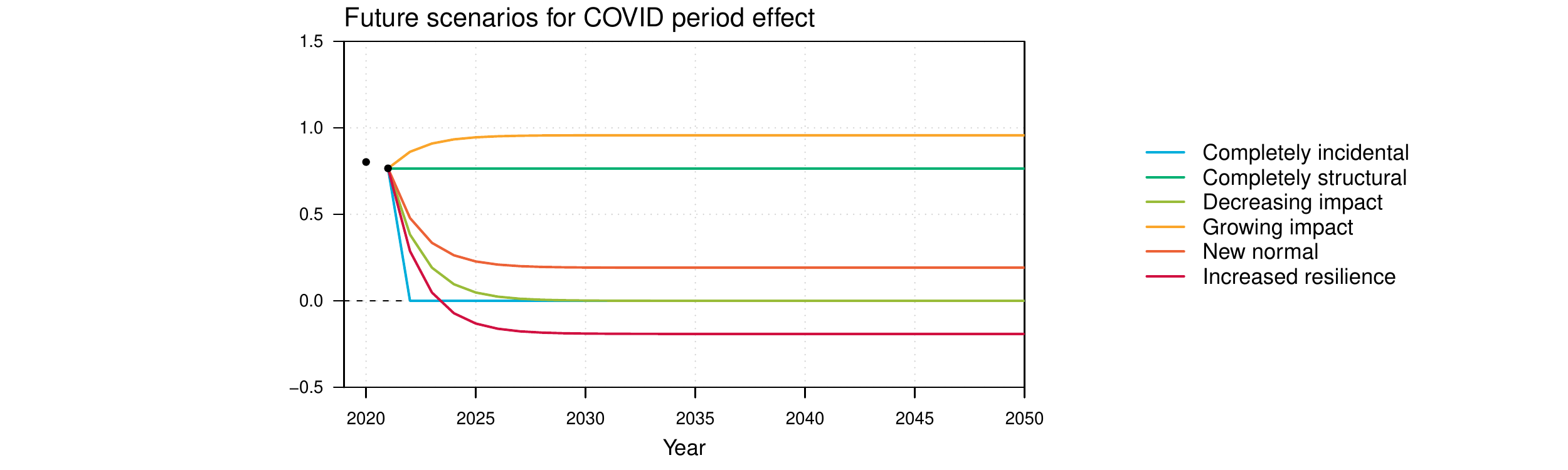}}
\caption{Projected COVID-19 period effects for Dutch males under various assumptions for future development of impact of COVID-19 on mortality.}
\label{fig:ForecastFrakK}
\end{figure}

We choose scenarios that can be written as:
$$
\begin{aligned}
  \ln \mu_{x,2021+h} &= \ln \mu_{x,2021+h}^{\text{pre-covid}} + \mathfrak{V}_x \mathfrak{X}_{2021+h} \\
  \mathfrak{X}_{2021+h} &= \mathfrak{X}_{\text{start}} \eta^h + (1-\eta^h)\mathfrak{X}_{\infty},
\end{aligned}
$$
for $0 \le \eta \le 1$. The value of the COVID-19 period effect in the limit, $\mathfrak{X}_{\infty}$, is chosen to be a multiple of $\mathfrak{X}_{2021}$ where the multiplicative factor is scenario-dependent. The parameter $\eta$ defines how fast convergence to this limit value takes place. This generic approach for generating COVID-19 scenarios is similar to that of \cite{ZhouLi2022} who estimate similar COVID-19 effects using a penalized quasi-likelihood approach. 
We define the following scenarios:
\begin{enumerate}
	\item \textbf{\texttt{Completely incidental}}: $\mathfrak{X}_{\text{start}}=\mathfrak{X}_{\infty}=0$.\newline
	The mortality forecast is completely based on the pre-COVID forecast.

	\item \textbf{\texttt{Completely structural}}: $\mathfrak{X}_{\text{start}}=\mathfrak{X}_{\infty}=\mathfrak{X}_{2021}$.\newline
	It is assumed that COVID-19 is a new cause of death that will remain permanently with an impact that does not change over time.

	\item \textbf{\texttt{Decreasing impact}}: $0<\eta<1$, $\mathfrak{X}_{\text{start}}=\mathfrak{X}_{2021}$, $\mathfrak{X}_{\infty}=0$.\newline
	It is assumed that the effect of COVID-19 converges to zero after a certain period, for example because herd immunity is (almost) reached in a population. 

	\item \textbf{\texttt{Growing impact}}: $0<\eta<1$, $\mathfrak{X}_{\text{start}}=\mathfrak{X}_{2021}$, $\mathfrak{X}_{\infty}=1.25\mathfrak{X}_{2021}$.\newline
	The impact grows, for example when new variants appear which are not impacted by current and new vaccines, or when other mitigating policies imposed by the government become less effective.
	
	\item \textbf{\texttt{New normal}}: $0<\eta<1$, $\mathfrak{X}_{\text{start}}=\mathfrak{X}_{2021}$, $\mathfrak{X}_{\infty}=0.25\mathfrak{X}_{2021}$.\newline
	The impact decreases to a constant but does not completely vanish, which results in a permanent effect that can be compared to other causes of death such as the flu.

	\item \textbf{\texttt{Increased resilience}}: $0<\eta<1$, $\mathfrak{X}_{\text{start}}=\mathfrak{X}_{2021}$, $\mathfrak{X}_{\infty}=-0.25\mathfrak{X}_{2021}$.\newline
	The impact gradually converges to a value below zero, for example because the population that survived the pandemic is stronger than the population before the pandemic.
\end{enumerate}
At the time of writing this paper, no information is available to make an informed decision for the value of $\eta$. We choose $\eta=0.5$ for illustration purposes. Given the other choices made for specifying the COVID-19 scenarios, the predicted period effects for Dutch males are shown in Figure~\ref{fig:ForecastFrakK}; the figure for females looks similar in case the same choices are made. 

We also need to specify how the COVID-19 age effect is defined for ages $x \notin \cal X$. Define $x_{\text{min}}$ and $x_{\text{max}}$ as the lowest respectively highest age included in the set of ages used for calibration, $\cal X$ (i.e. $x_{\rm min}=40$ and $x_{\rm max} = 95$ in the previous section). Figure~\ref{fig:NLD_parameters_age_selection} suggests that for lower ages the impact of COVID-19 on the level of mortality was negligible, and therefore these ages are excluded from calibration.  In line with that approach, we define $\mathfrak{V}_x=0$ for $x<x_{\text{min}}$. For higher ages, we observe that the estimated parameter $\mathfrak{V}_x$ {seems} relatively stable at higher ages. Based on that observation, we define $\mathfrak{V}_x=\mathfrak{V}_{x_{\text{max}}}$ for $x>x_{\text{max}}$.

\paragraph{Mortality predictions under various COVID-19 scenarios.} 

\begin{figure}[!h]
\centerline{\includegraphics[width=\textwidth, trim = 0cm 0.0cm 0cm 0cm]{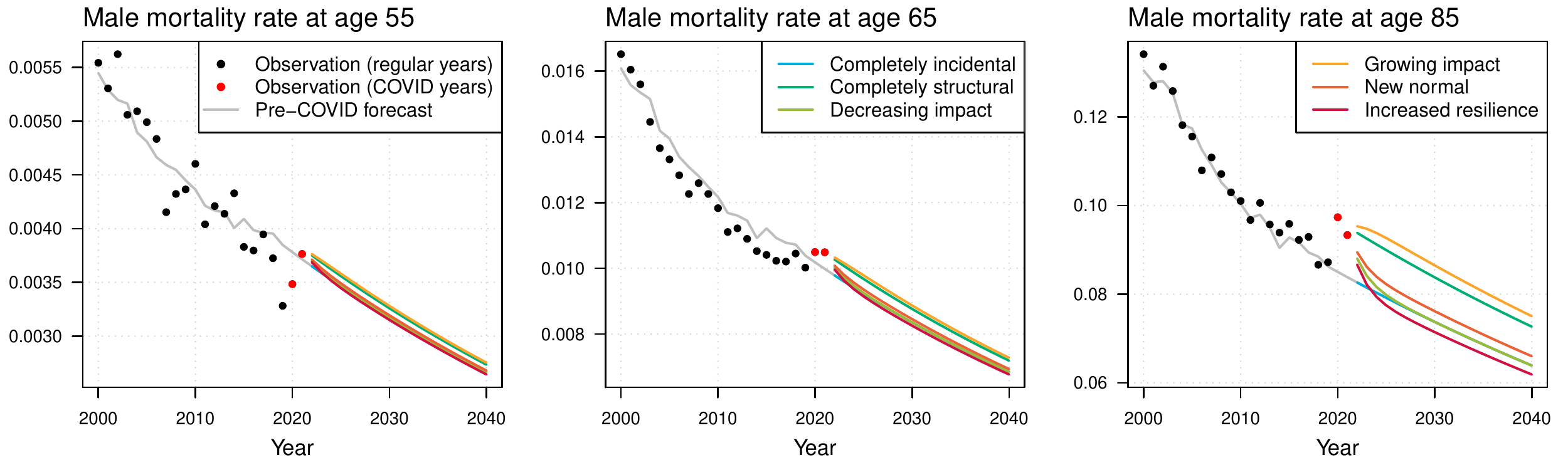}}
\caption{Projected mortality rates for Dutch males aged 55, 65 and 85 under various assumptions for future development of impact of COVID-19 on mortality.}
\label{fig:ForecastQx}
\end{figure}

For the various scenarios illustrated in Figure~\ref{fig:ForecastFrakK}, we have constructed the corresponding forecasts of mortality rates. Figure~\ref{fig:ForecastQx} shows the mortality forecasts for Dutch males at ages 55, 65 and 85. Though at age 55 mortality rates in 2020 and 2021 were not markedly different from the pre-COVID expectation, at age 65 the level of mortality was elevated and at age 85 it was substantially higher than expected. The pattern of the projected COVID-19 period effects from Figure~\ref{fig:ForecastFrakK} are clearly visible in the mortality forecasts. When constructing mortality rates, the COVID-19 period effects are multiplied with the COVID-19 age effects from Figure~\ref{fig:YearTransform} which explains why the forecast of mortality rates at age 55 are hardly affected by the COVID-19 period effect while at age 85 mortality in all scenarios (except for \texttt{Completely incidental}) deviates greatly from the pre-COVID forecast. For the scenarios \texttt{Decreasing impact}, \texttt{New normal} and \texttt{Increased resilience}, the impact on mortality diminishes quickly after a few years (given our choice for $\eta$), but for \texttt{Completely structural} and \texttt{Growing impact} the impact remains.

\begin{figure}[!h]
\centerline{\includegraphics[width=\textwidth, trim = 0cm 0.0cm 0cm 0cm]{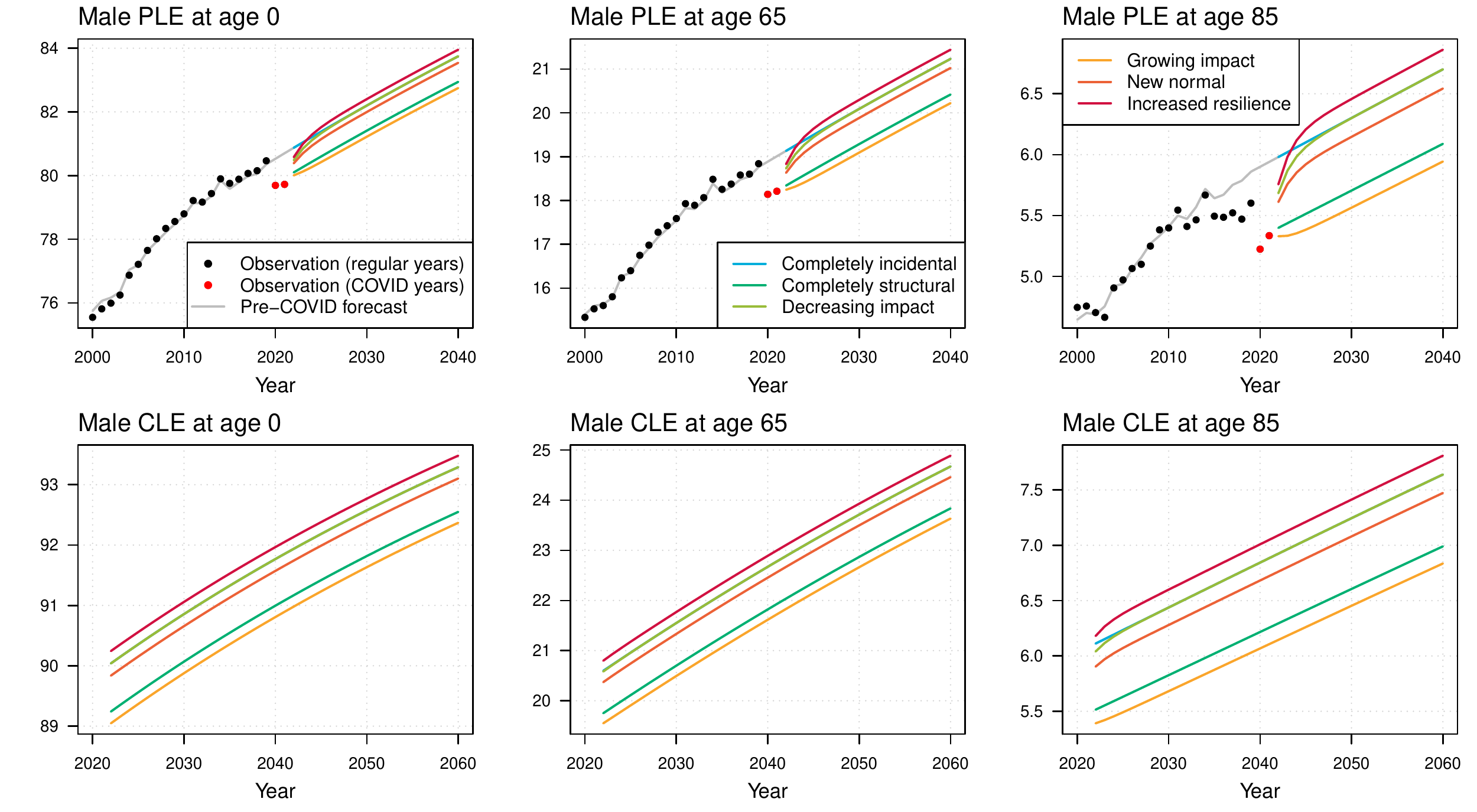}}
\caption{Projected period and cohort life expectancies for Dutch males at birth and at ages 65 and 85 under various assumptions for future development of impact of COVID-19 on mortality.}
\label{fig:ForecastLE}
\end{figure}

The impact on predicted period life expectancies and cohort life expectancies is shown in the top row respectively bottom row of Figure~\ref{fig:ForecastLE}. The period life expectancy at all ages is far below the pre-COVID expectation. Though the COVID-19 age effects are calibrated for the ages 40-95, the period life expectancy at age 0 is also affected, since mortality rates at all ages are used to compute this life expectancy. The gap between the most positive scenario (\texttt{Increased resilience}) and the most negative scenario (\texttt{Growing impact}) increases to approximately one year around 2040 for the ages shown. This impact remains constant further in the future once the COVID-19 period effect has converged to its limit value, and all remaining future developments are driven by the original Li-Lee model.

The pre-COVID forecast and the scenarios \texttt{Completely incidental} and \texttt{Decreasing impact} result in comparable projected cohort life expectancies.
While in the predicted period life expectancies differences between scenarios grow over time, in predicted cohort life expectancies the difference between scenarios at the start of the projection is close to the difference in the limit.

\section{Conclusions}\label{sec:Conclusion}

In this paper, we introduce a model to quantify weekly deviations from expected levels of mortality during a pandemic. The model 
{adds an extra layer to} the  two-layer Li-Lee model by including an additional seasonal effect to capture regular seasonal patterns and an additional age and week effect to measure deviations from pre-pandemic weekly mortality expectations. 

{We apply our model to data from Belgium, France, Germany and Great Britain using mortality data from the \textit{Short Term Mortality Fluctuations} database. There are differences between countries in the extent to which mortality at different ages is affected by COVID-19 and how COVID-19 affected mortality through time. Yet, there are also clear similarities since in all countries periods of high excess mortality are followed by periods of lower or no excess mortality.}

The application of the model requires the availability of exposures which often are not available {on a weekly basis} and  {these must therefore be approximated.} 
Useful insights on the development of mortality through a year can be obtained through a Composition Data (CoDa) analysis  {that can be} performed using weekly death counts only.

Most sources of weekly mortality observations provide data by gender and by age groups. Our sensitivity analysis shows that such data can be used to accurately monitor the development of the pandemic through time, which is represented by the COVID-19 week effect. However, the COVID-19 age effects are inaccurate in case of cohort effects exist in population sizes, and use of weekly observations by individual ages is therefore recommended. 


The future COVID-19 scenarios analyzed in this paper all assume convergence to pre-COVID long term improvement rates.
At this stage, there is insufficient data and information available to determine whether, and if so how long term mortality improvement rates should be adjusted.
This paper provides no solution for this problem, and this is a problem that is likely to challenge the imagination of demographers and actuaries for years to come.

\bigskip

\noindent{\bf Acknowledgment.}\
Certain parts of the approach proposed in this paper have been used in a model  to generate a prognosis for future survival probabilities  that was recently developed for the Royal Dutch Actuarial Association.
The authors gratefully acknowledge all comments from members of the
Association's 
Committee and Working Group on Mortality Research: Wies de Boer, Friso Cuijpers, Corn\'e van Iersel, Marieke Klein, Hans de Mik, Erica  Slagter, Janinke Tol, Erik Tornij,  Raymond Waucomont, Wouter van Wel, Menno van Wijk, Marco van der Winden and Kim Wittekoek. We also thank  Statistics Netherlands, and in particular Lenny Stoeldraijer, for helping us to obtain the required datasets.

\newpage

\bibliographystyle{plain}
\bibliography{RefsvBMVImpactPandemic021}

\newpage
\appendix

\section{Selection of ages}\label{sec:selection_of_ages}

In Figure~\ref{fig:NLD_parameters_seasonal} we observed that for ages below 40 the COVID-19 age effects are erratic. Since the age effects do not seem to capture a systematic COVID-19 effect at the younger ages, it might be better to exclude those ages from calibration of the model. The COVID-19 age effects are more stable from age 40 onward, though between ages 40 and 60 the estimated effect is close to zero.

\begin{figure}[!h]
\centerline{\includegraphics[width=\textwidth, trim = 0cm 0.0cm 0cm 0cm]{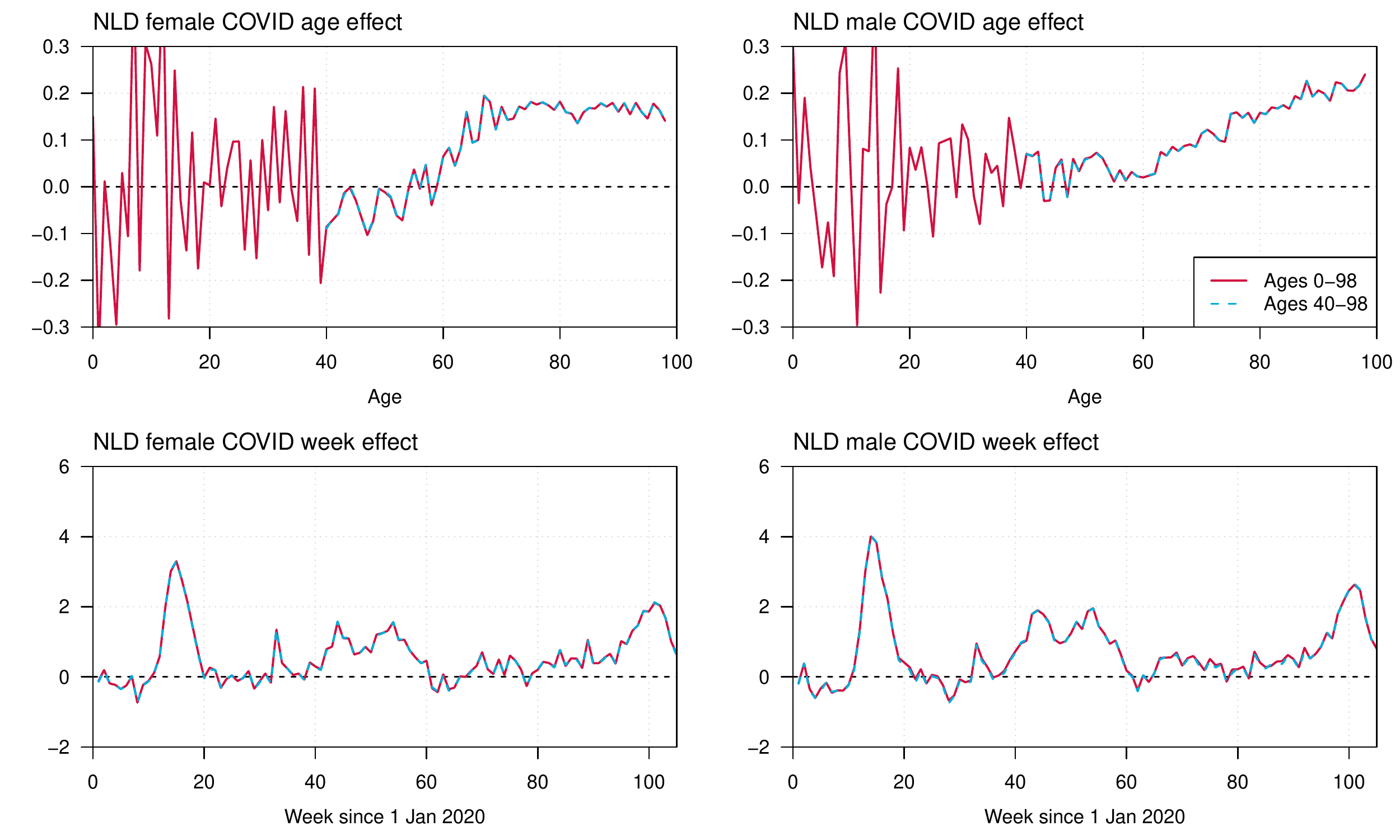}}
\caption{COVID-19 parameters estimated using ages 0-98 (dashed line) and ages 40-98 (solid line).}
\label{fig:NLD_parameters_age_selection}
\end{figure}

For interpretation of parameters and reliability of outcomes it is desirable that only structural effects are analyzed; noise due to low exposures should preferably be excluded.  {Therefore, we compare parameter estimates if we use different age ranges for calibration of the COVID-19 model. Figure~\ref{fig:NLD_parameters_age_selection} shows the estimated parameters for females and males using ages 0-98 (dashed line) and ages 40-98 (solid line) for calibration, where the former are renormalized such that the norms over the age range 40-98 are equal in both cases. This allows a comparison of the parameter estimates.}

From the parameter estimates in Figure~\ref{fig:NLD_parameters_age_selection} we conclude that the COVID-19 week parameters and age parameters for higher ages are hardly affected when including information from younger ages. However, the COVID-19 age effects for ages below 40 years are erratic, and therefore we exclude these ages from further analyses.

\newpage
\section{Additional calibration results based on STMF data}\label{sec:STMF_all_ages}

\begin{figure}[!h]
\centerline{\includegraphics[width=\textwidth, trim = 0cm 0.5cm 0cm 0cm]{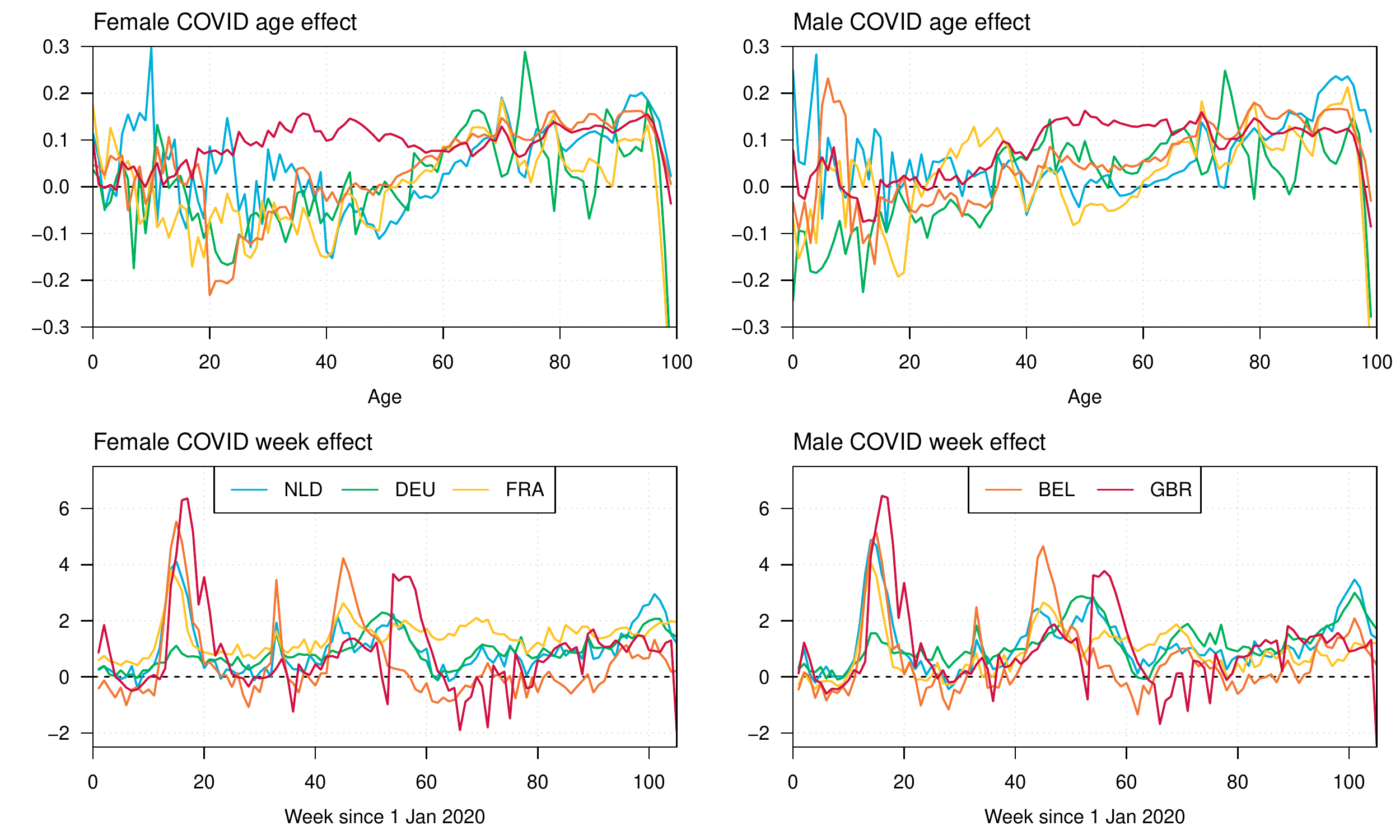}}
\caption{Estimated COVID-19 parameters for various countries using ages 0 up to and including 99.}
\label{fig:FiveCountriesAllAges}
\end{figure}

\end{document}